\documentclass[preprint]{aastex}
\newcommand \msol{M$_{\odot}$}

\newfont{\rten}{cmr10}

\begin{document}

%\normalsize

%\title{Model Spectral Energy Distributions of Circumstellar Debris Disks. II. Belt of Planetesimals with Interior Giant Planets}
\title{Signatures of Planets in Spatially Unresolved Debris Disks}
 
\author{Amaya Moro-Mart\'{\i}n\altaffilmark{1,2}, Sebastian Wolf\altaffilmark{2,3} \& Renu Malhotra\altaffilmark{4}}

\email{amaya@as.arizona.edu, swolf@astro.caltech.edu, renu@lpl.arizona.edu} 

\altaffiltext{1}{Steward Observatory, University of Arizona,
933 N. Cherry Avenue, Tucson, AZ 85721, USA}

\altaffiltext{2}{Max-Planck-Institut fur Astronomie, K\"onigstuhl 17, 69117 Heidelberg, Germany} 

\altaffiltext{3}{Division of Physics, Mathematics, and Astronomy, California Institute of Technology, Pasadena, CA 91125, USA}

\altaffiltext{4}{Department of Planetary Sciences, University of Arizona,
1629 E. University Boulevard, Tucson, AZ 85721, USA}

\begin{abstract}
Main sequence stars are commonly surrounded by debris disks, composed
of cold dust continuously replenished by a reservoir of undetected 
dust-producing planetesimals. In a planetary system with a belt of 
planetesimals (like the Solar System's Kuiper Belt) and one or more 
interior giant planets, the trapping of dust particles in the mean motion 
resonances with the planets can create structure in the dust disk, as the particles
accumulate at certain semimajor axes.  Sufficiently massive planets may also
scatter and eject dust particles out of a planetary system, creating a dust depleted
region inside the orbit of the planet. In anticipation of future observations of 
spatially unresolved 
debris disks with the $\it{Spitzer}$ $\it{Space}$ $\it{Telescope}$, we are interested 
in studying how the structure carved by planets affects the shape of the disk's spectral 
energy distribution (SED), and consequently if the SED can be used to infer the presence of 
planets. We numerically calculate the equilibrium spatial density 
distributions and SEDs of dust disks originated by a belt of 
planetesimals in the presence of interior giant planets in different planetary 
configurations, and for a  representative sample of chemical compositions. 
The dynamical models are necessary to estimate the enhancement of particles 
near the mean motion resonances with the planets, and to determine how 
many particles drift inside the planet's orbit. Based on the SEDs and 
predicted $\it{Spitzer}$ colors we discuss what types of planetary systems 
can be distinguishable from one another and the main parameter 
degeneracies in the model SEDs.
\end{abstract}


\section{Introduction}

Long before planets were discovered by the Doppler technique 
in the mid 90's, there was indirect evidence of planet building 
from observations of  $\it{debris~disks}$ around several main 
sequence stars (Backman \& Paresce~\citeyear{back93} and reference therein). 
Stars harboring debris disks are too old to have remnants of the 
primordial disk from which the star itself once formed.
The timescale of dust grain removal due to Poynting-Robertson (P-R) 
drag for a solar type star is t$_{PR}$$\sim$400$\times$R$^2$/$\beta$ years $\sim$ 10$^5$ years, 
where R is the grain distance to the central star in AU, and $\beta$ is the 
ratio of the radiation pressure force to the gravitational force (in the range 0--0.5).
Grain removal due to radiation pressure is much faster, as  the particles escape quickly 
on hyperbolic orbits. Grain removal timescales are therefore much shorter than the 
age of main sequence stars, $>$10$^7$ years, indicating that these dust disks are
not primordial. These Solar System-sized disks of micron-sized grains are 
thought to be the result of mutual collisions between asteroid-like 
bodies or the evaporation of comets close to the star (Backman \& Paresce \citeyear{back93}).

In the standard scenario, debris disks are generated 
mainly at early times when planetesimals are forming and colliding frequently;
this period would coincide with the heavy bombardment in the early Solar System.
In agreement with this scenario, far-infrared surveys with the $\it{Infrared}$ 
$\it{Space}$ $\it{Observatory}$ ($\it{ISO}$) indicate 
that disk detection drops abruptly at $\sim$0.4 Gyr (Habing et al. \citeyear{habi01}) and that 
the mass decline in the disks is proportional to t$^{-2}$ between stellar 
ages of 10 Myr and 1 Gyr (Spangler et al. \citeyear{span01}).  The processes responsible for 
the clearing of dust are stellar winds, radiation pressure, sublimation,
collisions and gravitational scattering by giant planets.
However, Greaves \& Wyatt (\citeyear{grea03}) found recently that a small number of stars
with an age of a few Gyr do have disks, in disagreement with the standard
scenario. They claim that Habing et al.~(\citeyear{habi01}) 
and Spangler et al.~(\citeyear{span01}) results are
biased toward younger ages, as they were preferentially detecting A (younger) 
stars. These new results indicate that debris disk duration is $\sim$0.5 Gyr, and may 
occur at any time during the main sequence, and that the disk mass decline in time 
is slow, not steeper than t$^{-1/2}$. It is suggested that the disks have ``on''
and ``off'' stages with large differences in dust mass. 
Furthermore, recent observations by the $\it{Spitzer}$ $\it{Space}$ $\it{Telescope}$
(Rieke et al.~\citeyear{riek05} and 
Gorlova et al.~\citeyear{gorl04}) suggest 
that in some cases the debris disk phenomena may be the result of stochastic 
catastrophic collisional events, rather than a continuous generation of dust for
long periods of time.

The observation of debris disks indicates that planetesimal formation is 
indeed a common by-product of the star formation process. One would expect, 
therefore, that some stars harbor both giant planets and extended 
emission, as is the case of the Sun. Submillimeter surveys have been 
designed to verify this assumption (Greaves et al.~\citeyear{grea04}), but the 
results are inconclusive due to the low sensitivity of the observations 
(down to a dust mass limit of 0.02 M$_{Earth}$), and the youth of some of
the stars (with difficult accurate radial velocity measurements). 
This situation is dramatically changing thanks to the high sensitivity of 
the $\it{Spitzer}$ MIPS instrument. A preview of a large $\it{Spitzer}$/MIPS GTO 
program confirms that out of 26 FGK field stars known to have planets by radial 
velocity studies, 6 show 70$\mu$m excess at 3-$\sigma$ confidence level, implying 
the presence of cool material ($<$100 K) located beyond 10 AU 
(Beichman et al.~\citeyear{beic05}). These stars, with a median age of 4 Gyr, 
are the first to be identified as having both well-confirmed planetary systems 
and well-confirmed IR excesses (Beichman et al.~\citeyear{beic05}). 
Additionally, there are high-resolution images of few debris disks that show 
the presence of density structure (Wilner et al.~\citeyear{wiln02} and 
Greaves et al.~\citeyear{grea98}). Dynamical models have shown that planets can sculpt 
the disks, creating gaps, arcs, rings, warps and clumps of dust 
(Roques et al~\citeyear{roqu94}; Liou \& Zook \citeyear{liou99}; Mouillet et al.~\citeyear{moui97}; 
Wyatt et al~\citeyear{wyat99}; Moro-Mart\'{\i}n \& Malhotra~\citeyear{ama02} and 
Kuchner \& Holman \citeyear{kuch03}), and therefore, it is suggested that these 
images confirm that debris disks and long-period planets coexist 
(Ozernoy et al.~\citeyear{ozer00}; Quillen \& Thorndike~\citeyear{quil02} and 
Wilner et al.~\citeyear{wiln02}).

We all know what is the fundamental question that missions like $\it{Terrestrial}$
$\it{Planet}$ $\it{Finder}$ will try to address: are there other potential sites for Life, 
far beyond our Earth? The study of debris disks can tell us where to look, and where not to look.
The reason why this is the case is the following: debris disk structure is sensitive 
to the presence of long-period planets, complementing a parameter space not covered 
by radial velocity and transient surveys, and an understanding of the orbits of long 
period planets is fundamental for the study of the stability of orbits in the habitable 
zone, where terrestrial planets could form and survive. 
In the future, the $\it{Atacama}$ $\it{Large}$ $\it{Millimeter}$ $\it{Array}$ ($\it{ALMA}$) 
will be able to image the dust in debris disks with an order of magnitude higher 
spatial resolution (10 milliarcseconds) than the $\it{Very}$ $\it{Large}$ $\it{Array}$ 
and the $\it{Hubble}$ $\it{Space}$ $\it{Telescope}$, in systems which are more than an 
order of magnitude fainter; i.e. it will be able to search for analogs of the Kuiper Belt 
dust disk. Careful modeling of the dust density distributions will be essential to interpret 
these high spatial resolution observational data in terms of planetary architecture. 
In the more immediate future, $\it{Spitzer}$ will carry out spectrophotometric 
observations of hundreds of circumstellar disks that will likely be spatially 
unresolved. This is why we are interested in studying how the dust density structure 
affects the shape of the disk spectral energy distribution (SED) and consequently 
if the SED can be used to infer the presence of planets.

A first general approach to the simulation of SEDs of debris disk 
systems using analytic dust density distributions has been 
undertaken by Wolf \& Hillenbrand (\citeyear{wolf03}; hereafter WH03).
This study made clear that the SED analysis strongly depends on the assumed 
density distribution, in particular of the smallest grain population. 
In contrast to the former approach by WH03, however, the dust density distribution of 
a debris disk should not be chosen $\it{a~priori}$ because it cannot be defined 
independently from the SED of the embedded star or the dust grain properties (grain 
size distribution, density and optical constants). In this 
study we use a self-consistent combination of existing numerical tools for the 
simulation of debris disk dust density distributions (that take into account 
the interplay between the central star SED, the grain properties and the dust dynamics), 
and the radiative transfer simulations in WH03 for the calculation of their emergent SEDs.

We study hypothetical debris disks originating from a belt of planetesimals [uniformly
distributed from 35 to 50 AU; similar to the Kuiper Belt (KB)]
and evolving under the effect of gravitational perturbation from interior giant planets 
in various planetary configurations.
In Moro-Mart\'{\i}n \& Malhotra (\citeyear{ama02} and \citeyear{ama03}) 
we described how in the Solar System, the trapping of particles 
in mean motion resonances (MMRs) with the giant planets can create 
structure in the KB dust disk as the dust particles accumulate at certain 
semimajor axes. We found that for the Solar System planetary configuration,
the azimuthal structure of the dust disk is not predictable in detail 
(with simulations of a small number N of particles, N$\sim$100) because it 
depends sensitively on the times of residence in the various resonances and these are highly 
variable and unpredictable. After careful analysis we concluded that even 
though the particle dynamics is chaotic, our method could robustly 
estimate the equilibrium radial density distribution of dust.
We found that the combination of radiation forces and planetary 
perturbations causes the dust disk to be depleted inward of Saturn's orbit and spread
outward beyond the KB source region, and the particle size distribution to flatten. 

As a complement to WH03, in this paper we investigate to what extent these 
planet-induced changes in the radial spatial density distribution and the 
particle size distribution affect the dust disk SED, and how these effects 
might be exploited to infer the presence of giant planets in 
$\it{spatially~unresolved}$ debris disks. In $\S$ 2, we calculate the 
spatial density distribution of dust grains of different sizes, 
corresponding to different $\beta$ values. The dynamical evolution of 
the dust particle, and therefore the dust density distribution, 
depend only on the parameter $\beta$, which is a function of the 
grain size and composition and the stellar spectrum (assumed to be solar). 
The resulting three-dimensional density distribution is transformed into 
a one-dimensional radial density distribution, which is sufficient to calculate the disk 
SED because we assume that the disk is optically thin, and therefore the 
temperature distribution of the grains depends only on the distance
to the central star. In $\S$ 3, we select a representative sample of 
chemical compositions based on debris disk spectroscopic observations, 
including Fe-rich and Fe-poor silicates (crystalline and amorphous olivine and 
pyroxene) and carbonaceous materials. For each chemical composition, using 
laboratory optical constants and Mie theory, we calculate the grain
radiation pressure coefficients averaged over the stellar spectrum. This 
allow us to find the correspondence between $\beta$ and the particle 
radius. Once the particle size is known for each $\beta$ value and selected chemical 
composition, in $\S$ 4 we calculate, using the same one-dimensional radiative 
transfer code as in WH03, the emitted dust SED (plus the stellar scattered light) 
for each single particle size, single composition disk. We later weight and 
combine these SEDs in order to consider certain particle size distributions. 
Because this work is part of the $\it{Spitzer}$ FEPS Legacy project, 
which is focused on the detection and characterization of debris disks 
around F,G and K nearby stars, we use a solar type star as the central 
heating source and a distance of 50 pc (the median distance of the FEPS targets).  
A disk mass of 10$^{-10}$~\msol~ has been assumed to ensure that the disk is optically 
thin to stellar radiation, even along the mid plane, and that mutual grain 
collisions are not important. We further assume that the disks have little 
or no gas, so that the dust dynamics is controlled by gravitation and radiation 
forces only.  The modeling does not include mutual grain collisions and gas drag (i.e.
the systems under consideration are old optically thin disks); quick sublimation 
of icy fraction (a rapid mass loss can cause the grain's orbit to become more 
eccentric); grain erosion due to sputtering by solar wind particles; and Lorentz 
forces due to interplanetary magnetic fields. 
For a more detailed description of the applied numerical model
and an estimate of the limitations mentioned above we refer to  
Moro-Mart\'{\i}n \& Malhotra (\citeyear{ama02} and \citeyear{ama03}).

A schematic diagram explaining modeling approach is shown in Figure 1.
The goal is to investigate how the presence of planets affects the 
debris disk's SED by comparing systems with planets (of different
masses and different semimajor axis) and systems without planets, and to 
derive and analyze the parameter degeneracies in the model SEDs.

\section{Dust Spatial Density Distributions}
A detailed description of the dynamical models used to calculate the 
dust spatial density distributions, the numerical algorithm 
used to integrate the equations of motion, and the uncertainties inherent 
in the prediction of structure, owing to the chaotic dynamics of dust orbital 
evolution, were given by Moro-Mart\'{\i}n \& Malhotra (\citeyear{ama02}). 
Here, we briefly overview the main ideas. 

We follow numerically, from source to sink, the evolution of sets of 100 
dust particles from an outer belt of planetesimals similar to the 
KB under the combined effects of solar gravity, solar radiation 
pressure, Poynting-Robertson (P-R) and solar wind drag, and the gravitational 
forces of the planets. We study the following planetary configurations: 
\begin{enumerate}
\item The Solar System with 7 planets (excluding Mercury and Pluto). 
The parent bodies of the dust particles are assumed to be uniformly distributed 
in orbits with semimajor axis between 35 and 50 AU, eccentricities 
such that the periastron distances are between 35 and 50 AU (i.e. between
0 and 0.3), and inclinations between 0$^\circ$ and 17$^\circ$, in approximate 
accord with current estimates of the orbital distribution of KB objects (KBOs; 
Malhotra et al.~\citeyear{malh00}). 
\item A system with the same distribution of parent bodies as above but
without planets.
\item Nine single-planet systems with a planet mass of 1, 3 and 10 M$_{Jup}$
in a circular orbit with semimajor axis of 1, 5, and 30 AU. The parent bodies are distributed
in orbits with semimajor axis between 35 and 50 AU, eccentricities 
between 0 and 0.05 and inclinations between 0 and 3$^\circ$,
to account for the fact that a thinner planetesimal disk may be more
realistic when a single giant planet is present. [In this study we found that
the difference between the ``thick'' and the ``thin'' planetesimal disks is negligible].
For the models with a single planet at 30 AU, we ignored the dust particles originated 
from the 30 planetesimals that lie between 35--40 AU [i.e. inside 
the 3:2 MMR]. This is because due to multiple close encounters with 
the planet, we do not expect to have planetesimals in stable orbits in that region. 
\item A system with the same distribution of parent bodies as above but
without planets.
\end{enumerate}

For the first two cases (KB-like disk with and without Solar System planets), 
we run 17 sets of 100 particles each, corresponding to 17 different particle sizes, 
with $\beta$ values ranging from 0.00156 to 0.4, distributed to get a uniform 
logarithmic sampling in particle size (see Figure 5). For the rest of the systems 
(i.e. the ``thinner'' planetesimal disk with and without single planets), 
we run a subset of nine  $\beta$ values, also ranging from 0.00156 to 0.4. 

We assume that the dust is generated from a constant grinding down of planetesimals 
due to mutual collisions or collisions with interstellar grains, adopting a constant dust
production rate throughout the planetesimal belt (between 35 AU and 50 AU). 
The sinks of dust included in our numerical simulations are
ejection into unbound orbits, accretion onto the planets, and 
orbital decay to less than r$_{min}$, where r$_{min}$ = 0.5 AU (astrocentric distance) 
for all models, except for those where the single planet is at 1 AU, for which we use 
r$_{min}$ = 0.1 AU instead. 
%[This minimum distance is determined by the time step of the 
%integrator so that $\it{dt}$ is at least 6 times the period of a massless 
%particle orbiting at r$_{min}$.] 
Assuming that the dust production rate is in equilibrium with the loss rate, and
the dust particle dynamics is ergodic (i.e. the time-weighting reflects
the spatial density), we can obtain equilibrium density distributions 
by recording the positions of these 100 particles at equal time intervals
(every 1000 years); then transforming the particles' coordinates into 
a reference frame rotating with the planet dominating the structure (Neptune);
and finally treating each position as an individual particle, i.e. 
accumulating all the rotated particles' coordinates over the total lifetime of the sample
particles. This leads to a three-dimensional time-weighted equilibrium 
density distributions that is ``resampled'' into a logarithmic 
one-dimensional radial grid which is the input for the radiative transfer 
code. 

\subsection{Radial Density Distributions: Output from the Dynamical Models}
Figure 2 shows some of the resulting surface density distributions of dust.
The main features are the following:
\begin{enumerate} 
\item When no planets are present the dust density distribution is flat, as
expected for a collisionless system with grains in circular orbits 
(Briggs~\citeyear{brig62}), and no large particles ($\beta$$<$0.5) are found 
at distances larger than  the apoastron of the parent bodies.
But when planets are present, the surface density distribution deviates
for a flat profile (see below) and gravitational scattering of dust by the 
giant planets is able to extend the disk beyond the boundaries set by 
radiation effects alone. 
\item Depletion of dust inside the planet's orbit due to gravitational scattering 
by the planet. In the Solar System, depletion takes place in the inner 
10 AU from gravitational scattering by Jupiter and Saturn.
Inner cavities have been also inferred to exist
in systems like $\beta$ Pic (20 AU), HR 4796A (30-50 AU), 
$\epsilon$ Eri (50 AU), Vega (80 AU) and Fomalhaut (125 AU) 
(Dent et al.~\citeyear{dent00}; Greaves, Mannings \& Holland
2000; Wilner et al.~\citeyear{wiln02} and Holland et al.~\citeyear{holl03}).
Some of these systems are highly collisional, a regime where our modeling
approach is not valid. The cavities, however, are possibly created by gravitational 
scattering with an inner planet.
\item Enhanced dust density in a ring outside the planet's orbit. This is produced by
the trapping of particles in exterior MMRs with the planet. In the 
Solar System, the ring is between 35--50 AU and the resonant planet is Neptune. 
The trapping into MMRs can clearly be seen in the ``equilibrium'' semimajor
axis distributions shown in Figure 3 (for the single-planet models) and 
Figure 10 in Moro-Mart\'{\i}n \& Malhotra (\citeyear{ama02}) (for the Solar System models).
\item The structure is more pronounced for larger particle sizes (smaller $\beta$)
because the trapping in MMRs is more efficient when the drag forces are small.
The boundary of the disk is less steep for smaller particles (larger $\beta$)
compared to larger particles; this is because immediately
after release from their parent bodies the orbits of the former 
are more strongly affected by radiation pressure, which tends to
increase their eccentricity and semimajor axis.
\end{enumerate} 
The surface density distributions in Figure 2 show in some cases
scattering at small astrocentric distances. In others cases, the particles 
drifting inward do not follow a flat surface density (instead it rises steeply). 
The presence of planets may explain part of these features, but from 
experience we know that some of this ``noise'' owes to our use of 
a logarithmic radial sampling to allow higher spatial resolution near 
the central star (where the high grain temperature implies a 
strong contribution to the SED). Where the radial shells are very small, 
a particle crosses many radial grids before its position is recorded; 
this, together with the fact that we are modeling the dynamical 
evolution of a small number of test particles (N$\sim$100), produces numerical 
``noise'' due to small number statistics. For this reason, as described in 
$\S$ 2.2, the radial density distributions that are used as input for the 
radiative transfer code will not take into account the numerical results 
at small astrocentric distances.

The depletion factor inside the planet's orbit is the percentage of particles that
are ejected from the system, relative to the total number of particles. Because we are modeling 
sets of only 100 particles, and jovian-mass planets eject a significant fraction of these, 
the number of particles that drift inward is usually small and is subject to some 
uncertainty. We have several KB models with the same or similar initial conditions
whose results indicate that a conservative estimate of the uncertainty in the depletion 
factor is $\sim$10\% of the initial number of particles. Depletion factors are 
larger than 90\% for the majority of the single-planet systems studied 
(and for most $\beta$ values), except for 1M$_{Jup}$ at 1 AU and 5 AU for
$\beta$$>$0.025, for which depletion factors are $\sim$60\%--80\% and  $\sim$50\%--80\%, 
respectively. This means that except for these two cases, the uncertainty due to the 
small number of particles studied makes the depletion factors obtained consistent with 
having an empty hole (i.e. 90\%$\pm$10\% consistent with 100\% depletion). 

\subsection{Radial Density Distributions: Input for the Radiative Transfer Models}
The radial density distributions are uncertain interior to the planet's orbit
to a 10\% level. To account for this uncertainty, and to estimate the contribution
of the particles trapped in the MMRs to the SED of the disk, we will 
calculate and compare the SEDs that arise from three different types of 
surface density distributions (see Figure 4):
\begin{enumerate} 
\item $\it{Empty~Gap}$ models: the surface density distribution accounts 
for the trapping of particles in the MMRs with the planet 
and the total depletion of particles interior to the planet's orbit, 
i.e. we assume that the 
``gap''\footnote{In this paper a ``gap'' is an inner 
hole interior to the planet's orbit, not an angular depletion zone around the 
planet's orbit.} is empty (100\% depletion). The astrocentric distance of this 
``gap'' is determined by the radius at which the surface density from 
the numerical results decreases by more than 90\%.
Outside the gap, the surface density distribution follows the results 
from the dynamical models (large dots). 
\item $\it{Partial~Gap}$ models: same as above but with the inner hole being
90\% depleted in dust with respect to the disk without planets (instead of
being totally empty, or 100\% depletion). In this case, we are extrapolating the 
surface density of dust from a distance near the planet's orbit down to 0.01 AU, 
to account for the sublimation distance of the larger silicate grains. 
The detailed calculation of the sublimation distance is done by the radiative transfer 
code, and depends on the grain radius and chemical composition.
The extrapolation is done assuming a flat surface density distribution, 
expected for a collisionless system with grains in circular orbits.
The caveat is that when planets are present the dust grains that drift inward may
have a non-zero eccentricity, so the surface density will not be exactly flat. 
\item $\it{Analytical~Gap}$ models: the surface density distribution consists 
on a simple square profile, following the flat density distribution of the 
disk without planets, with an empty gap at the planet's position (i.e. it
does not account for particles trapped in the MMRs). 
\end{enumerate} 

The comparison of the SEDs that arise from the 
model with an $\it{empty~gap}$ and the model with a $\it{partial~gap}$ 
can teach us if the dynamical models are  $\it{sufficient}$ to distinguish the presence 
of planets of masses ranging from 1--10 M$_{Jup}$ and semimajor axis of between 1--30 AU.  
If we find that the SEDs arising from these two models are significantly different, 
the number of particles studied (N$\sim$100) would not be sufficient, as it does not 
allow us to distinguish between an empty gap and a partial gap with 10\% of particles left.
The comparison of the SEDs that arise from the $\it{empty~gap}$ and the $\it{analytical~gap}$ 
models can teach us if the dynamical models are  $\it{necessary}$, or whether it is adequate to assume a 
flat surface density distribution with a clean gap inward of the planet's location (ignoring the 
accumulation of dust particles in the MMRs).

The calculation of these radial density distributions is the most CPU demanding 
step, but it is independent of the rest of the steps outlined
in the Introduction because the dynamical models and the resulting radial 
density profiles depend only on the parameter $\beta$. 
It is only $\it{a~posteriori}$ that we find the 
relationship between the particle size and composition and the $\beta$ value, 
with a single $\beta$ value corresponding to several combinations of grain
size and composition (see Figure 5). Because of this degeneracy, 
our scheme allows enough flexibility to efficiently explore other grain chemistries 
without the need of recomputing the radial density profiles. 

\section{Correspondence between $\beta$ and Particle Size}
The quantity $\beta$ is the dimensionless ratio of 
the radiation pressure force and the gravitational force. 
For spherical grains and a solar type star, 
$\beta$=5.7 $\times$ 10$^{-5}$ Q$_{pr}$/($\rho${\it a}), where $\rho$ 
and {\it a} are the density and radius of the grain in cgs units 
(Burns, Lamy \& Soter~\citeyear{burn79}), and 
Q$_{pr}$ is the radiation pressure coefficient.
Before using the radial density distributions as input for the
radiative transfer code in WH03, we need to find the correspondence between 
the value of $\beta$ and the particle size, and this depends on the 
grain chemical composition. 
Detailed chemical analysis of young, massive circumstellar disks around T Tauri 
and Herbig Ae/Be stars show that the dust consists mainly of two distinct species: 
a silicate component and a carbonaceous component (Savage \& Mathis \citeyear{sava79}; 
Draine \& Lee \citeyear{drai84} and Malfait et al. \citeyear{malf00}). 
$\it{ISO}$ spectroscopy of Herbig Ae/Be 
stars revealed several other chemical components, such as Fe 
(broad, weak emission feature at 24 $\mu$m; see Henning \& Stognienko \citeyear{henn96}), 
FeO (broad emission features at 2125 $\mu$m; see Henning et al. \citeyear{henn95}), and/or 
FeS (see Henning \& Stognienko \citeyear{henn96}) and H$_2$O ice (broad emission features 
between 40 and 80 $\mu$m; Warren \citeyear{warr84}).
The SEDs of debris disks measured so far, however, do not allow one to perform 
a comparably detailed chemical analysis. Mid-IR spectroscopy of $\beta$-Pic can 
be explained with dust grains composed of 55\% amorphous olivine, 35\% amorphous pyroxen 
and 10\% crystalline olivine (see e.g. Pantin, Lagage \& Artymowicz~\citeyear{pant97}). 
The zodiacal light silicate feature can be matched by a mixture of amorphous
forsterite/olivine, dirty crystalline olivine and a hydrous silicate
(Reach et al.~\citeyear{reac03}). Based on these spectroscopic observations,
the following chemical compositions were selected:
MgSiO$_3$ and Mg$_{0.6}$Fe$_{0.4}$SiO$_3$ (Fe-poor and Fe-rich pyroxene), 
MgFeSiO$_4$ and Mg$_{1.9}$Fe$_{0.1}$SiO$_4$ (amorphous and crystalline 
olivine), and C400 and C1000 [400 K carbon modification (graphite-poor) 
and 1000 K carbon modification (graphite-rich), respectively].
This is a subset of the compositions studied in WH03; 
we refer to this paper for a justification of this selection and a description
of the optical properties of these silicate and carbon species.
The SEDs presented in this study correspond to disks composed of only 
one grain type (of the six listed above). A mixture of grains is 
more realistic, with the grain composition probably varying from 
source to source (see e.g. Reach et al.~\citeyear{reac03}). 
Because we are assuming that the disks are optically thin,
so that the grains do not interact with each other, we can account for
different mixtures by linearly combining the single grain composition SEDs 
into a final SED, using different contributing factors for each of the 
grain chemical compositions. This step will be taken in the future when using 
our models to simulate $\it{Spitzer}$ observations. 

For each selected chemistry, using Mie scattering theory
and assuming that the grains are homogeneous spheres, we compute the
grain optical parameters, needed to calculate dust absorption, reemission and 
scattering of radiation. We obtain Q$_{pr}$ as a function of wavelength for 
a large number of particle sizes.
%\footnote{The code for calculation of the Mie scattering coefficients
%is available at http://mc.caltech.edu/$\sim$swolf/miex-web/miex.htm}. 
The quantity Q$_{pr}$ is a function of the grain complex refractive
indexes ($\it{n}$, $\it{k}$), the grain radius, and the wavelength of the 
incoming radiation. 
The refractive index for the silicates and carbonaceous
materials are taken from  Dorschner et al. (\citeyear{dors95}) and 
J\"{a}ger et al. (\citeyear{jage98}), 
respectively\footnote{The complex refractive indexes are available at http://www.astro.uni-jena.de/Laboratory/Database/odata.html.}.
We then obtain the average of Q$_{pr}$ integrated over the solar 
spectrum (Labs \& Neckel~\citeyear{labs68}). This average is
used to calculate the value of $\beta$, for each particular dust 
chemistry under consideration, and for a large number of particle
sizes (see Figure 5). Finally, we select the particle size whose 
$\beta$ is closer to the $\beta$-value adopted in the dynamical models.
One important feature to notice from Figure 5 is that, for a given particle 
size, the value of $\beta$ corresponding to carbonaceous and Fe-rich silicate grains 
is larger than that of Fe-poor silicates, because
the former have a very high absorptive efficiency in the wavelength
range on which the star emits. This is important because a small change 
in the abundance of carbonaceous and Fe-rich silicate 
material can make a very significant change in the level of the continuum emission.

The parameter study in WH03 showed that the shape of the SED is 
affected by the relative number of small grains, which is determined
by the minimum ($\it{a}_{min}$) and maximum ($\it{a}_{max}$) grain size, and
by the index ($\it{q}$) of the power law size distribution ($\it {n(a)da=n_0a^{-q}}$): 
an increase of net flux and the prominent emission features occur when $\it{a}_{min}$ is 
decreased and $\it{q}$ is increased. 
%For $\beta$ Pic, MIR observations indicate that there is a substantial amount of
%grains $<$ 10 $\mu$m inside 20 AU (Weinberger et al.~\citeyear{wein03}), 
%while others suggest $\it{a}_{min}$$\approx$0.1 $\mu$m (Pantin, Lagage \& Artymowicz~\citeyear{pant97}).
%For Fomalhaut, submillimeter observations indicate
%$\it{a}_{min}$$\approx$0.7 $\mu$m (Wyatt \& Dent~\citeyear{wyat02}). 
In this paper $\it{a}_{min}$=0.5--1.3 $\mu$m
(depending on composition), and is determined by the condition $\beta$=0.5, corresponding to particles
that are forced into hyperbolic orbits as soon as they are released from
their parent bodies. If the parent bodies' orbits have eccentricity $\it{e}$, 
ejection occurs for $\beta >$ 0.5(1-{\it e}) and  $\beta >$ 0.5(1+{\it e}) for particle 
release at periastron and apoastron, respectively.
We are therefore implicitly assuming that radiation pressure is the only 
process responsible for the minimum grain size, but in practice $\it{a}_{min}$ 
is also affected by collisional processes.
The maximum grain size in our simulations is limited by the CPU time, as particles with very
small $\beta$s (0.00156 is our minimum value) have a very slow dynamical
evolution. Depending on the chemical composition chosen, $\it{a}_{max}$=53--244 
$\mu$m. Debris disks certainly contain larger ``dust'' particles, up
to planetesimal size, but in the wavelength range considered grains larger 
than $\sim$1 mm will not contribute significantly to the SED,
and the missing grains in the 53--240 $\mu$m to 1 mm range only add an
almost featureless continuum, as indicated in the study by WH03.

\section{Spectral Energy Distributions}
Once the particle size is known for each $\beta$ value and selected chemical 
composition, we use the surface density distributions as input for a 
radiative transfer code that calculates the emitted dust SED 
(plus the stellar scattered light). 
As in WH03, we assume that the disk is optically thin:
only scattering, absorption and reemission of stellar
radiation by dust grains are taken into account, 
neglecting multiple scattering and radiation 
and dust heating due to dust reemission. 
The dynamical models are only valid in a density regime 
that corresponds to optically thin disks. 
For the central star we use the solar SED published by 
Labs \& Neckel (\citeyear{labs68}), 0.2--100 $\mu$m, extended by a blackbody SED
(T=5800 K) beyond 100 $\mu$m.  
The dust reemission and scattering are calculated at 500 logarithmically 
equidistantly distributed wavelengths between 5--340 $\mu$m (which includes
the wavelengths covered by $\it{Spitzer}$).
We assume a disk mass of 10$^{-10}$~\msol~and a distance of 50 pc.
Note that our models (with and without planets) contain the same amount of disk mass.  
We are interested in studying how the structure created by the planets affects the 
shape of the SED, independent of the dust production rate. However, 
planetary perturbations can affect the dust production rate, 
possibly leading to more massive dust disks.  This effect is not taken into 
account in our models, but will be considered in the future.

Figure 6 shows the SEDs that result from the Solar System models.
Each color corresponds to the SED that arise from a single 
particle size disk (we only show three $\beta$ values of the 17 computed).
Each panel corresponds to a particular grain chemical composition.
In Figure 7 we compare the emission arising from the different 
compositions (keeping the particle size approximately constant).
In agreement with WH03, the most important features shown in these
figures are: 
\begin{enumerate} 
\item Emission is stronger for Fe-rich silicates 
(Mg$_{0.6}$Fe$_{0.4}$SiO$_3$ and MgFeSiO$_4$) compared to 
Fe-poor (MgSiO$_3$ and Mg$_{1.9}$Fe$_{0.1}$SiO$_4$). 
This is due to the strong dependence of the UV-to-NIR 
absorption efficiency on the Fe content, which leads to higher 
grain temperatures at a given distance from the star as the Fe content 
increases.
\item Similarly to Fe-rich silicates, carbonaceous grains
also lead to stronger but featureless emission (mainly adding
a continuum). 
\item The emission peak shifts to longer wavelengths as the 
particle size increases (or $\beta$ decreases). This is due to the fact 
that the turnover point beyond which the absorption efficiency
decreases continuously increases with grain size.
\item The clearing of dust from the inner 10 AU results 
in a loss of warm dust and is responsible for the decrease in the 
NIR/MIR region (compared to the case when no planets are present). 
The slight shift in the emission peaks indicate that some of 
this NIR/MIR emission is radiated at longer wavelengths (as the 
cleared particles are located further away from the star), but this
is a very small effect because once the particles are set on 
hyperbolic orbits after their last gravitational encounter with 
the giant planet, they leave the system very quickly without  
contributing significantly to the emission. 
The net flux decreases because a larger fraction of the grains
are further away from the star, so the fraction of 
stellar photons that the grains can absorb and later 
re-emit is diminished.
\end{enumerate} 

Similarly, but not shown here, we have calculated single particle size
and single composition SEDs for the other planetary systems studied 
(i.e. the nine single-planet models and the system without planets). 
These SEDs show features similar to those described above for the 
Solar System.

As Figure 2 shows, the structure of the dust disk is significantly different 
depending on the particle size under consideration. The structure is more pronounced 
for larger particle sizes because the trapping in resonances is more efficient when 
the drag forces are small. However, it is expected that the dust production 
processes will favor the generation of small particles. The modeling of debris 
disk structure and SED should therefore take into consideration an appropriate 
range of particle sizes that can later be weighted and combined to emulate
a particle size distribution.
In Moro-Mart\'{\i}n \& Malhotra (\citeyear{ama03}) we estimated the 
radial distribution of KB dust from our dynamical models and the KB 
dust production rate estimates from  Landgraf et al. (\citeyear{land02}). 
We showed that the dust particle size distribution in space is significantly 
changed from its distribution at production, due to the combined effects of 
radiation forces and the perturbations of the planets. 
Radiation forces alone change the differential size distribution from the
(assumed) initial power law of index $q=3.5$ at production (corresponding to
a fragmentation power law), to a shallower power law with $q\approx2.5$, 
valid at distances smaller than the aphelion of the parent bodies. 
Planetary perturbations further affect the power law index because
larger particles are more easily trapped in MMRs.
With these results in mind, the single particle size and single composition 
SEDs for each planetary system (like those shown in Figure 6 for the Solar System) 
are weighted and combined in such a way that particle size distribution throughout
the disk (not necessarily where the dust is generated) follows power laws of 
indexes $\it{q}$=2.5, 3.0 and 3.5 (i.e each SED  corresponds to a dust mass of 
10$^{-10}$~\msol~and is weighted in such a way that the total mass in each 
particle size bin follows a power law).
Note that when collisional processes are considered in detail, 
the particle strengths are size-dependent, leading to a size-dependent 
$\it{q}$; and because the particle growth/collision processes 
depends on the radial distance from the star, $\it{q}$ will also be a function
of radius. The trapping of particles in MMRs with the planets also
adds a radial dependency to the power law index. 
Here, we will ignore these effects and consider a single power law to 
describe the particle size distribution at all distances from the star
and for all particle sizes.

The top panel of Figure 8 shows that the flux is higher, specially in the NIR/MIR range,  
and the spectral features are more pronounced when the particle size 
distribution is steeper (i.e. there is a larger fraction of smaller 
grains). This is because small grains achieve higher temperatures.
%Question: why feature is more pronounced?
In WH03 it was found that because of this effect, the presence of a gap,
and consequently the removal of warm grains, lead to a more pronounced 
decrease of flux in the NIR/MIR range when a larger fraction of smaller 
grains were present (steeper power law).
Figure 8 (bottom) suggest exactly the opposite trend, as
the ratio of the SEDs that arise from a system with a planet (using 
the empty gap models) to that from a system without a planet (i.e. without 
a ``gap'') is smaller for smaller power law indexes. The discrepancy between 
this result and the one in WH03 
arises from the difference between the analytic density 
distributions used in WH03, and the density distributions 
used here; the latter showing a large difference between the system with 
planets and the system without planets, which is more pronounced for larger 
grains (smaller $\beta$) than for smaller grains (larger $\beta$; 
see Figure 2). 
%This is due to the fact that: (1) the trapping in MMRs 
%is more efficient when the P-R drag force is small (small $\beta$), 
%and (2)  gravitational scattering extends the disk beyond the boundaries 
%set by radiation effects alone; the difference between the outer 
%disk radius in the system with planets and the system without planets 
%is much larger for larger particles (smaller $\beta$). 
This means that the difference between the mean disk temperatures in the system with
planets and that of the system without planets is more pronounced when large 
particles are dominant, leading to more distinct differences in their 
corresponding SEDs. This illustrates the importance of combining  numerical tools for 
the simulation of debris disk structure with a detailed radiative 
transfer code for the calculation of their emergent SEDs.
%In WH03, the use of an analytic radial density distribution (a power law 
%with a gap) could lead to the conclusion that the difference 
%between the SEDs for a system with planets (with a ``gap'') and a system
%with no planets (without a ``gap'') is larger when the size
%distribution is steep. But we have seen here that when you account for
%the full dynamics of the dust, this tendency is reversed, and the 
%difference is larger when large particles dominate.

The SEDs that result after combining the different grain sizes are shown 
in Figure 9. Each panel corresponds to a different 
planetary system (indicated at the top). Figure 10 shows the ratios of the SEDs 
that arise from a system with a planet (using the empty gap 
models) to that from a system without a planet (i.e. without a ``gap''). In the following, we 
refer to this ratio as F$_{planet}$/F$_{no~planet}$, where F is the flux arising
from the dust disk. Similarly, we show F$_{partial~gap}$/F$_{empty~gap}$ 
and F$_{empty~gap}$/F$_{analytical~gap}$.
The SED depends on the grain chemical composition. For example, Figure 9 
shows that for carbonaceous and Fe-rich silicates grains, the minimum 
in the SED of a dust disk with Solar System-like planets is at 
$\lambda$$<$8 $\mu$m, while for Fe-poor silicate grains the minimum 
shifts to longer wavelengths, $\lambda$=10--25 $\mu$m. A similar 
effect is also found for the other single-planet systems studied. Similarly, 
the minimum of F$_{planet}$/F$_{no~planet}$ occurs at $\lambda$$<$10 $\mu$m for 
carbonaceous and Fe-rich silicates grains, and $\sim$ 15 $\mu$m for 
Fe-poor silicate grains, i.e., the wavelength range where the difference 
between the SED arising from a disk with Solar System-like planets and that of a 
disk without planets is the largest depends on the chemical composition
of the grains. As it can be seen from Figure 10,
in many cases the largest difference between F$_{planet}$ and F$_{no~planet}$ 
occurs at wavelengths where the photospheric emission from the star dominates, 
making the photospheric subtraction critical in the analysis of observed 
SEDs in terms of planetary architectures.

Figure 10 shows that for a planet at 1 AU, the differences 
between the ``empty gap'' and the ``analytical gap'' models
are large, with F$_{empty~gap}$/F$_{analytical~gap}$ up to 100 for 1M$_{Jup}$, 
and 30 for 10M$_{Jup}$ (dashed lines). 
This is due the fact that the ``analytical gap'' models follow a square profile, 
while the dynamical models contain a large number of particles accumulated in 
the MMRs with the planet (see Figure 2 and note that scale is logarithmic). 
These particles, being at small astrocentric distances, are hot and contribute 
very significantly to the SED. We can conclude that dynamical simulations 
are necessary to model the SEDs of debris disks in the presence of 
planets at small semimajor axis (hot Jupiters), because the enhancement 
of particles at the MMRs dominates the emission. A consequence of this 
is that it should be possible to distinguish observationally between a simple 
square profile for the surface density of the dust disk, as that created by a 
stellar wind or by the interaction of the dust grains with ambient gas, from the 
surface density created by the dynamical interactions with a massive planet.  

Figure 10 also shows that for a planet at 1 AU, the ratio F$_{planet}$/F$_{no~planet}$
(solid lines) is greater than 1 for $\lambda$=8--60 $\mu$m for carbonaceous and Fe-rich 
silicate grains, and $\lambda$=20--80 $\mu$m for Fe-poor silicate grains. 
Even though the disk with a planet has an inner hole, it can be up to 3 times 
brighter than the disk without a planet. This is because the particles accumulated 
in the MMRs contribute importantly to the SED.
The system with 10M$_{Jup}$ at 1 AU, however, is not significantly brighter than the system
without planet.
This does not mean that the analytical square profiles are sufficient for 
10M$_{Jup}$ at 1 AU, because as we saw above, F$_{empty~gap}$/F$_{analytical~gap}$ $\sim$ 30.
The ratio  F$_{planet}$/F$_{no~planet}$ is very close to 1, for $\lambda$$>$80 $\mu$m 
for 1M$_{Jup}$ at 1 AU, and $\lambda$$>$24 $\mu$m for 10M$_{Jup}$ at 1 AU. In 
this wavelengths ranges, either the effect of the particles accumulated in the resonances
is not important, or their effect on the SED is balanced by the depletion of hot grains
close to the star. 

The ``partial gap'' and ``empty gap'' models are very similar for 1M$_{Jup}$ 
at 1 AU (F$_{partial~gap}$/F$_{empty~gap}$ $\sim$ 1; dotted lines in Figure 10).
For a 10M$_{Jup}$ planet at 1 AU, F$_{partial~gap}$/F$_{empty~gap}$ $\sim$ 3, for $\lambda$$<$20 $\mu$m. 
But in this case, our models indicate that the number of particles that drift inward 
is probably 0, with an uncertainty smaller than 10\%, which means that 
F$_{partial~gap}$/F$_{empty~gap}$ is probably overestimating the uncertainties 
in the prediction of the SED.
We conclude from the small F$_{partial~gap}$/F$_{empty~gap}$ values 
that the uncertainties in our dynamical simulations, due to the small number of particles 
studied, do not affect the modeling of the SEDs arising from dust disks with 
a planet at 1 AU. In other words, the number of particles in our simulations 
(N$\sim$100) is sufficient to model these systems with close-in planets.

As we saw before, for 1M$_{Jup}$ at 1 AU the accumulation of dust grains in the 
MMRs with the planet can increase the flux up to a factor of 100 compared to the flux 
arising from a disk with a simple square profile. 
For 1M$_{Jup}$ at 5 AU, the maximum F$_{empty~gap}$/F$_{analytical~gap}$
decreases from $\sim$100 to $\sim$30, and therefore the 
accumulation into the MMRs is not as important as at 1 AU  (the particles 
are colder and their contribution is less dominant). 
But the difference between the ``partial gap'' and ``empty gap'' models
is more pronounced at 5 AU (F$_{partial~gap}$/F$_{empty~gap}$ $\sim$3--10 
for 1M$_{Jup}$ and 10--30 for 10M$_{Jup}$) than at 1 AU 
(F$_{partial~gap}$/F$_{empty~gap}$ $\sim$1 for 1M$_{Jup}$ and 3 for 10M$_{Jup}$). 
[The number of particles that 
drift inward for 10M$_{Jup}$ at 5 AU is probably 0 and not subject to the 
10\% uncertainty, so the factor of 10--30 is probably overestimated.]
We conclude that the dynamical models are necessary to study the SEDs arising 
from systems with  planets of 1--10M$_{Jup}$ at 5 AU, because by not considering
the particles accumulated in the MMRs, the SED can be underestimated by a factor of
30. But unlike the 1 AU models, the number of particles that we have used in our 
simulations (N$\sim$100) is not sufficient because a 10\% uncertainty in depletion 
factor yields to a factor of 3--10 and 10--30 (overestimated) in flux 
for 1 and 10M$_{Jup}$, respectively.

The models at 30 AU represent a system with a narrow ring of dust producing
planetesimals just outside the planet's orbit. If a fraction of the dust 
particles drift inward, even if it is small, the shape of the SED is similar 
to the one arising from a system without planets. This is because like in 
the models without planets, there is no structure for a wide range of astrocentric distances.
If the particles are held back (by trapping in resonances and gravitational 
scattering), the SED shows a large deficit in the mid-IR flux (whose wavelength 
depend on the location of the ring) that makes it very distinct from the SED 
from a system without planets (see large minimum for F$_{planet}$/F$_{no~planet}$
in Figure 10).

The ``empty gap'' and ``analytical gap'' models are very similar when the planet is 
at 30 AU, i.e. the accumulation of particles in the MMRs do not dominate the shape
of the SED for planets at this distance. However, the ``partial gap'' models are very 
different from the ``empty gap'' models, with the 10\% uncertainty in the 
number of particles that drift inward leading to a factor of 100--3000 difference in flux
(F$_{partial~gap}$/F$_{empty~gap}$ $\sim$ 100, for 1M$_{Jup}$ and 
F$_{partial~gap}$/F$_{empty~gap}$ $\sim$ 1000--3000, for 10M$_{Jup}$).
Because the SED is very sensitive to the number of particles that drift inward, 
this number needs to be determined precisely by dynamical models; the large factors indicate
that the number of particles in our simulations (N$\sim$100) is not sufficient
to study systems with planets of 1--10M$_{Jup}$ at 30 AU.

The ratio F$_{planet}$/F$_{no~planet}$ can reach 0.3 for 1M$_{Jup}$ at 1 AU, 
and 0.1 for 10M$_{Jup}$ at 1 AU. This decrease is due to the fact that for 
the more massive planet the gap is larger (r$_{gap}$$\sim$0.8 
AU for 1M$_{Jup}$, and~$\sim$ 1.6 AU for 10M$_{Jup}$; see Figure 4), and 
more empty (larger number of particles are ejected).
The different depletion factors explain why a disk with 1M$_{Jup}$ at 1 AU (5 AU) is brighter 
than a disk with 10M$_{Jup}$ at 1 AU (5 AU) by a factor of 3--10 (for $\lambda$$<$10 $\mu$m for 
carbonaceous and Fe-rich silicate grains; and $\lambda$ between 10--24 $\mu$m 
for Fe-poor silicate grains).
The different gap radius makes the disk with a 1M$_{Jup}$ at 1AU be 100 (3000) times brighter 
than a disk with 1M$_{Jup}$ at 5 AU (30 AU) (for $\lambda$$<$24 $\mu$m, if the grains 
have carbonaceous and Fe-rich silicate composition;  and $\lambda$$<$50 $\mu$m, if
they have Fe-poor silicate composition). 

\section{Predicted $\it{Spitzer}$ Broadband Colors}

We have calculated expected $\it{Spitzer}$ broadband colors.
In principle, the SEDs in Figure 9 should contain all the information given by 
the color-color diagrams. However, the 
advantage of these diagrams is that one can compare easily the 
results arising from many different models, allowing to explore more 
efficiently the parameter degeneracies in the model SEDs. In particular,
we are interested in exploring the effects of planet mass and
location, particle size distribution and composition, and the depletion
factor inside the gap cleared by the planet.

The central bandpass wavelengths are  
4.51 $\mu$m and 7.98 $\mu$m (IRAC), and 23.68 $\mu$m, 71.42 $\mu$m and 
155.9 $\mu$m (MIPS)\footnote{Spectral responses are available at http://ssc.spitzer.caltech.edu/irac/spectral$\_$response.html and http://ssc.spitzer.caltech.edu/mips/spectral$\_$response.html.}. 
In addition, we have integrated the SEDs using square profiles centered at 13.2 $\mu$m and 
32.5 $\mu$m, with widths of 1.6 $\mu$m and 5.0 $\mu$m (for the 
IRS observations). These widths are chosen to avoid
the bad segments of the IRS instrument and the long-wavelength tail of 
the 10 $\mu$m SiO feature. Figure 11 shows five different combinations 
of color-color diagrams (one per row). Each panel corresponds to a different grain 
chemical composition (only three are shown).
The different symbols correspond to different planetary systems: the 
symbol shape indicates the planet semimajor axis and the 
the symbol color indicates the planet mass, the power law index for the 
particle size distribution, and whether the gap is empty or partially filled.

The types of planetary systems that can be distinguishable from one another based
on their $\it{Spitzer}$ colors are listed in Table 1. The main results are the following:
\begin{enumerate}
\item The colors considered here can be used to diagnose the location of the planet 
(in the 1--30 AU range) and the absence/presence of planets, if the gap cleared
by the planet is depleted by more than 90\%.
\item Except for one particular case, in general it is not possible to diagnose the 
mass of the planet (in the 1--10M$_{Jup}$ range) based on these colors.
\item If the disks are composed of carbonaceous grains (C400 and C1000), the different
planetary systems considered have indistinguishable $\it{Spitzer}$ colors, i.e. the
higher the carbonaceous content the more difficult it is to diagnose planetary systems
from their colors.
\item If instead of ``empty gap'' models we consider ``partial gap'' models, where 
10\% of the particles drift inward, most of the planetary systems in Table 1 become 
indistinguishable from one another. It is important to keep this in mind when
looking at the results in Table 1  involving planets at 5 or 30 AU, as in these cases
our dynamical models do not contain enough particles to determine precisely 
how many of them drift inward (so we are subject to a $\sim$10\% uncertainty).
\item In some cases (see Table 1), two planetary systems that are distinguishable 
for one index of the power law for the grain size distribution, are not 
distinguishable when using a different index ($\it{q}$=2.5 versus $\it{q}$=3.5). This,
together with the fact that different compositions yield different results (see 
e.g. how different are the colors for MgSiO$_{3}$ grains compared to the other 
compositions), complicates the analysis of the colors, as the particle size 
distribution and chemical composition are not known.
\item 24$\mu$m/32$\mu$m and 70$\mu$m/160$\mu$m colors are not useful to distinguish among 
the planetary systems considered in this study (1--10M$_{Jup}$ at 1--30AU).
\end{enumerate}

\section{SED Probability Distribution}
The previous analysis describes how the SED is affected by variations of
specific parameters (e.g. the presence/absence of planets, the chemical
composition of the dust, and the particle size distribution). Because in 
many cases these parameters are unknown, in this section we consider the 
degeneracy that is created by the entire suite of parameters.
In order to do that, we have selected the four radial density distributions shown in the 
upper left panel of Figure 2. They correspond to a disk with a 1M$_{Jup}$ planet 
located at 1 AU, and four different values of $\beta$ (or grain sizes): 
0.00156, 0.0125, 0.1 and 0.4. For each of one of these four 
distributions, we considered 21 different depletion factors: from 
80\% to 100\% (empty gap) in increments of 1\%. We then ``shifted''
each one of these 4$\times$21 distributions in the radial coordinate, 
to simulate the radial density distributions that would arise if the planet is 
located at a larger semimajor axis. We considered 36 planet locations
between 1 AU and 30 AU. This adds up to a total of 4$\times$21$\times$36 
different radial density distributions. Although these distributions
are not exactly the ones that would result from the dynamical models
(if we were to run 4$\times$21$\times$36 different disk models),
a comparison with the results at 5.2 AU and 30 AU shows that 
they are a reasonable approximation, as they include the main features: 
a range of dust depletion factors inside the orbit of the planet, and
accumulation of particles in the MMRs with the planet. 
For each one of these distributions, and selecting one grain mineralogy, we used
the radiative transfer code to calculate its SED.
The SEDs from the four different grain sizes (or $\beta$-values) are 
combined using 11 different power laws indexes (from 2.5 to 
3.5, in increments of 0.1), resulting in a total of 21$\times$36$\times$11 
SEDs. We repeated this procedure for the six different chemical
compositions considered in our study. The results are shown in 
Figures 12 and 13. 

If the dominant grain composition is known, Figure 12 indicates that, 
for the six compositions considered, the shape of the SED of a 
disk with embedded planets is very distinct from the SED of a disk
without planets. The main difference, as mentioned before, 
is the decrease of the near-mid-IR flux ($<$40 $\mu$m), due to 
the clearing of particles inside the planet's orbit. Observations
at long wavelengths ($>$40 $\mu$m) are important to determine the 
shape of the SED, so we can distinguish whether the decreased near-mid-IR
flux is due to a region depleted of dust, or to an overall less massive
disk.  However, if the dominant grain composition is unknown, Figure 13
indicates that it is not possible to distinguish between a disk
with planets and a disk without planets. This illustrates the 
importance of obtaining spectroscopy observations able to constrain the 
grain chemical composition of the dust. Regardless of the chemical
composition, there are two regions where the dispersion of the SEDs
is small ($\sim$ 1 order of magnitude in flux). As expected, one is
at the longest wavelengths ($>$250 $\mu$m), where the emission is 
dominated by the largest grains, and therefore the spectral features are not
as important (except for crystalline olivine); 
and the other region is at $\sim$60 $\mu$m (as was also 
pointed out by WH03). If the disk and dust grain 
properties are unknown, these two regions are the most diagnostic for 
the determination of the total disk mass. 

For each one of these SEDs we have calculated expected $\it{Spitzer}$ 
broadband colors. The results are shown in Figure 14. For carbonaceous grains, 
the colors from a disk with embedded planets are indistinguishable from those of a disk 
without planets (as was pointed out in $\S$5). The same happens 
for Fe-rich silicate grains, except for 70$\mu$m/160$\mu$m, where we could
marginally distinguish between a disk with planets and a disk without 
planets (but only for certain planetary configurations). However, if Fe-poor 
silicate grains dominate, Figure 14 indicates that the colors can indeed be used 
to diagnose the presence or absence of embedded planets, in particular,
8$\mu$m/13.2$\mu$m, 24$\mu$m/70$\mu$m  and 24$\mu$m/32$\mu$m.

\section{Conclusions and Future Work}
In anticipation of future observations of spatially unresolved debris disks with 
$\it{Spitzer}$, we are interested in studying how the structure carved by 
planets affects the shape of the disk's SED,
and consequently if the SED can be used to infer the presence of 
planets. We numerically calculate the 
equilibrium spatial density distributions of dust disks composed of different 
grain sizes, originated by a belt of planetesimals similar to the KB and
in the presence of interior giant planets in different planetary configurations. 
A radiative transfer code is used to generate their corresponding SED 
for a  representative sample of grain chemical compositions.
The goal is to find the main parameter degeneracies in the model SEDs
and the distinguishing characteristics between the SEDs of 
different planetary configurations.

In practice, the modeling of an observed SED is done by using 
simple analytical surface density distributions defined by parameters that 
can be varied to fit the observations. However, these parameters cannot be chosen
arbitrarily, independently from the SED of the embedded star or the dust 
grain properties. A self-consistent combination of numerical models for the 
simulation of debris disk dust density distributions (that take into account 
the interplay between the central star SED, the grain properties and the dust dynamics), 
and a radiative transfer code is needed for the calculation of the dust disk SED.
Our models indicate that for close-in planets (1AU), an important parameter to consider is the 
enhancement factor in a ring-like structure located outside the planet's orbit,
and related to the number of particles accumulated in the MMRs.
Hot Jupiters can trap dust particles in MMRs at small astrocentric 
distances. These particles are hot and can have an important contribution 
to the SED. Trapping in resonances can therefore make the disk to look 
brighter, facilitating its detection, but it also makes a disk with
an inner planet less distinguishable from a disk without planet, as 
the clearing of hot dust inside the planet's orbit (and its corresponding 
decrease of the near-mid-IR flux) is compensated by the trapping of particles in MMRs.
For planets at larger semimajor axis (5 AU and 30 AU), the important parameter 
to consider is the density drop interior to the planet's orbit, related
to the fraction of particles that are able to drift inward. These two parameters,
describing the density enhancement and the density drop,
depend on the mass and location of the planet, and can only be estimated 
using dynamical simulations. We conclude that: (1) dynamical models are necessary 
to study the SEDs arising from debris disk systems with embedded planets 
of 1--10M$_{Jup}$ at 1--30AU; and (2) the number of particles in the dynamical 
simulations presented here (N$\sim$100) is sufficient to study systems with 
planets at 1 AU, but for the study of planets at  5 AU and 30 AU we need 
to increase the number of particles in our simulations in order to improve the statistical 
uncertainty in the number of particles that drift inward to better than 10\%.

The SED of the dust disk depends on the grain properties (chemical composition, 
density and size distribution) and the mass and location of the perturbing planet. 
The SED of a debris disk with interior giant planets is fundamentally different 
from that of a disk without planets, the former showing a significant decrease 
of the near/mid-IR flux due to the clearing of dust inside the planet's orbit. 
The SED is particularly sensitive to the location of the planet, i.e. to 
the area interior to the planet's orbit that is depleted in dust (see Figure 15). 
However, there are some degeneracies that can complicate the interpretation 
of the SED in terms of planet location.
For example, the SED of a dust disk dominated by Fe-poor silicate grains has its minimum at 
wavelengths longer than those of a disk dominated by carbonaceous and Fe-rich 
silicate grains. Because the SED minimum also shifts to longer wavelengths 
when the gap radius increases (owing to a decrease in the mean temperature 
of the disk), we note that there might be a degeneracy between the dust grain chemical 
composition and the semimajor axis of the planet clearing the gap. For an example, 
notice the similarities in the shape of the SEDs arising from a dust disk with 
a 3 M$_{Jup}$ planet at 1 AU and dominated by MgSiO$_3$ grains, and 
a disk with a 3 M$_{Jup}$ planet at 30 AU and dominated by MgFeSiO$_4$ grains 
(see Figure 16).  This illustrates the importance of obtaining 
spectroscopy observations able to constrain the grain chemical composition, 
and/or high resolution images, able to spatially resolve the disk.

We saw that for planets at 5 and 30 AU, the difference in the SED arising 
from a disk with an empty gap interior to the planet's orbit, and a disk with
an inner hole that is 90\% depleted, is large. This means that the SED 
is very sensitive to the depletion factor inside the gap. Because this 
depletion factor depends largely on the planet mass, in principle one 
should be able to use SEDs to diagnose masses of planets at large 
astrocentric distances. To do that we would need to increase the number
of particles in our dynamical simulations. Based on the depletion factors
obtained from our models, we expect that the planetary masses that one 
would be able to study with this method would range from $>$1M$_{Nep}$ to
3M$_{Jup}$. Anything larger than 3M$_{Jup}$ would create an almost empty 
gap, being indistinguishable from one another. A 1M$_{Nep}$ planet at 
30 AU ejects $<$10\% of the particles, so the efect of a less massive planet
would probably be difficult to detect. At 1 AU, the SED of the disk seems to be 
more insensitive to the depletion factor (because as mentioned above, 
its effect on the SED is balanced by the particles in the MMRs). This means that
it would be difficult to diagnose the mass of a close-in 
planet based on the SED of the dust disk.

The gaps and azimuthal asymmetries observed in high resolution images of 
debris disks suggest that giant planets may be present in these systems. 
Because debris disk structure is sensitive to a wide range of planet 
semimajor axis, complementing a parameter space not covered by  
radial velocity and transit surveys (sensitive only to close-in planets),
the study of the disk structure can help us learn about the diversity 
of planetary systems. Even when spatially resolved images of the disk are 
not available, we have seen that its SED may contain the signatures of the 
underlying planets. The SED can therefore be a valuable tool for detecting
and even constraining the mass and location of the planet. 

Once interesting $\it{Spitzer}$ targets have been identified, the next step 
will be to obtain  high-sensitivity and high-spatial resolution images
in scattered light and/or thermal emission (using e.g. $\it{LBT}$, 
$\it{JWST}$, $\it{Sofia}$, $\it{ALMA}$ or $\it{Safire}$). 
Of particular interest are the longer wavelengths, 
where observations can constrain the amount of material further away from the planet, 
and where the emission of the larger dust particles, the ones that show more 
prominent structure, dominate. If one could obtain observations that
spatially resolve the disk, the dynamical models could allow us to 
locate the perturbing planet. Then we could compare the information 
derived from the SED alone to that derived from the resolved image. 
This is important for the understanding of the limitations of the 
characterization of planetary architectures based on spatially unresolved debris 
disks only. Also, by obtaining resolved images in one or more wavelength we
can break the degeneracy expected from the analysis of the disk SED. 
In anticipation of these spatially resolved observations, we have started working 
on the modelling of the brightness density distributions 
arising from debris disks in the presence of different planetary 
configurations. 

\begin{center} {\it Acknowledgments} \end{center}
We thank Hal Levison for providing the SKEEL computer code, and Dana Backman,
Martin Cohen, Lynne Hillenbrand, Jonathan Lunine, Michael Meyer, 
Alberto Noriega-Crespo and Steve Strom for useful discussion. We also 
want to thank the referee for her/his careful reading of the manuscript.
This work is part of the {\it Spitzer} FEPS Legacy project (http://feps.as.arizona.edu). 
We acknowledge NASA for research support (contracts 1224768 and 1407 
administered by JPL and grants NNG04GG63G, NAG5-11661 and NAG5-11645) and
IPAC and the $\it{Spitzer}$ Science Center for 
providing access to their facilities. S. Wolf was also supported through the 
DFG Emmy Noether grant WO 857/2-1.

\clearpage

\begin{deluxetable}{lccc}
\tablewidth{0pc}
\tablecaption{Planetary Systems with Distinct Colors}
\tablehead{
\colhead{Composition} &
\colhead{4$\mu$m/8$\mu$m} & 
\colhead{8$\mu$m/13$\mu$m} & 
\colhead{8$\mu$m/24$\mu$m}}
\startdata
MgSiO$_3$                    & 1,5,30AU---no pl     & 1Jup1AU---1Jup30AU\tablenotemark{a} & 30AU---no pl \\
                             &                      & 1Jup1AU---1Jup5AU\tablenotemark{a}  & 1,5AU---30AU \\ 
                             &                      & 1AU---30AU               &              \\
                             &                      & 5AU---nopl               &              \\
Mg$_{0.6}$Fe$_{0.4}$SiO$_3$  & 1AU---5AU\tablenotemark{a}      & 1,5AU,no pl---30AU\tablenotemark{a} &              \\
                             & 1AU---30AU           &                          &              \\
                             & 5AU---no pl\tablenotemark{a}    &                          &              \\
                             & 30AU---no pl         &                          &              \\
                             & 1Jup5---10Jup5\tablenotemark{b} &                          &              \\
MgFeSiO$_4$                  & 1AU---30AU\tablenotemark{b}     &                          &              \\
Mg$_{1.9}$Fe$_{0.1}$SiO$_4$\tablenotemark{a} &                 & 1AU---5,30AU\tablenotemark{a}       &              \\
                             &                      & 5,30AU---no pl\tablenotemark{a}     &              \\
\enddata
\tablenotetext{~}{\textrm List of planetary systems with distinct $\it{Spitzer}$ colors. 
Notation: ``1,5AU---30AU'' means that the models with the planet at 1 or 5 AU 
are distinguishable from the models with the planet at 30 AU. 1Jup1 is a 
1M$_{Jup}$ planet at 1 AU. All results refer to ``empty gap'' models.}
\tablenotetext{a}{only valid for $\it{q}$=2.5} 
\tablenotetext{b}{only valid for $\it{q}$=3.5}
\end{deluxetable}

\clearpage

\begin{deluxetable}{lcc}
\tablewidth{0pc}
\tablecaption{Planetary Systems with Distinct Colors}
\tablehead{
\colhead{Composition} &
\colhead{13$\mu$m/24$\mu$m} &
\colhead{24$\mu$m/70$\mu$m}}
\startdata
MgSiO$_3$                    & 1,5AU---30AU & 1Jup1---1Jup30\tablenotemark{b} \\
                             & 30AU---no pl &                      \\ 
MgFeSiO$_4$                  &              & 1Jup1---1Jup30\tablenotemark{b} \\
\enddata
\end{deluxetable}

\clearpage

\begin{figure}
\epsscale{1.0}
\plotone{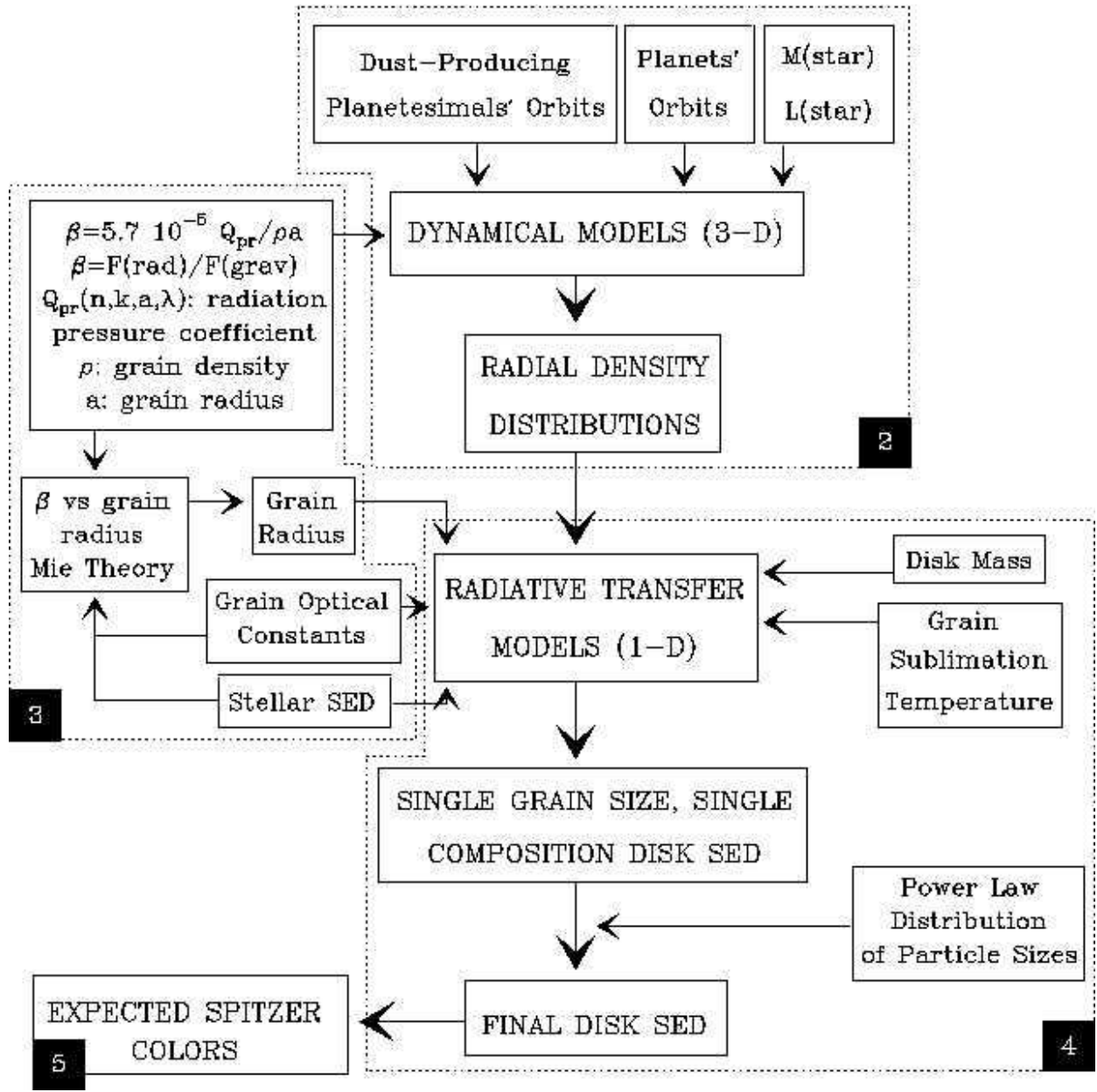}
\caption{Schematic diagram showing the steps of the modeling approach. 
The numbers inside the black squares indicate the relevant sections 
in this paper}
\end{figure}

\clearpage 

\begin{figure}
\epsscale{1.0}
\plotone{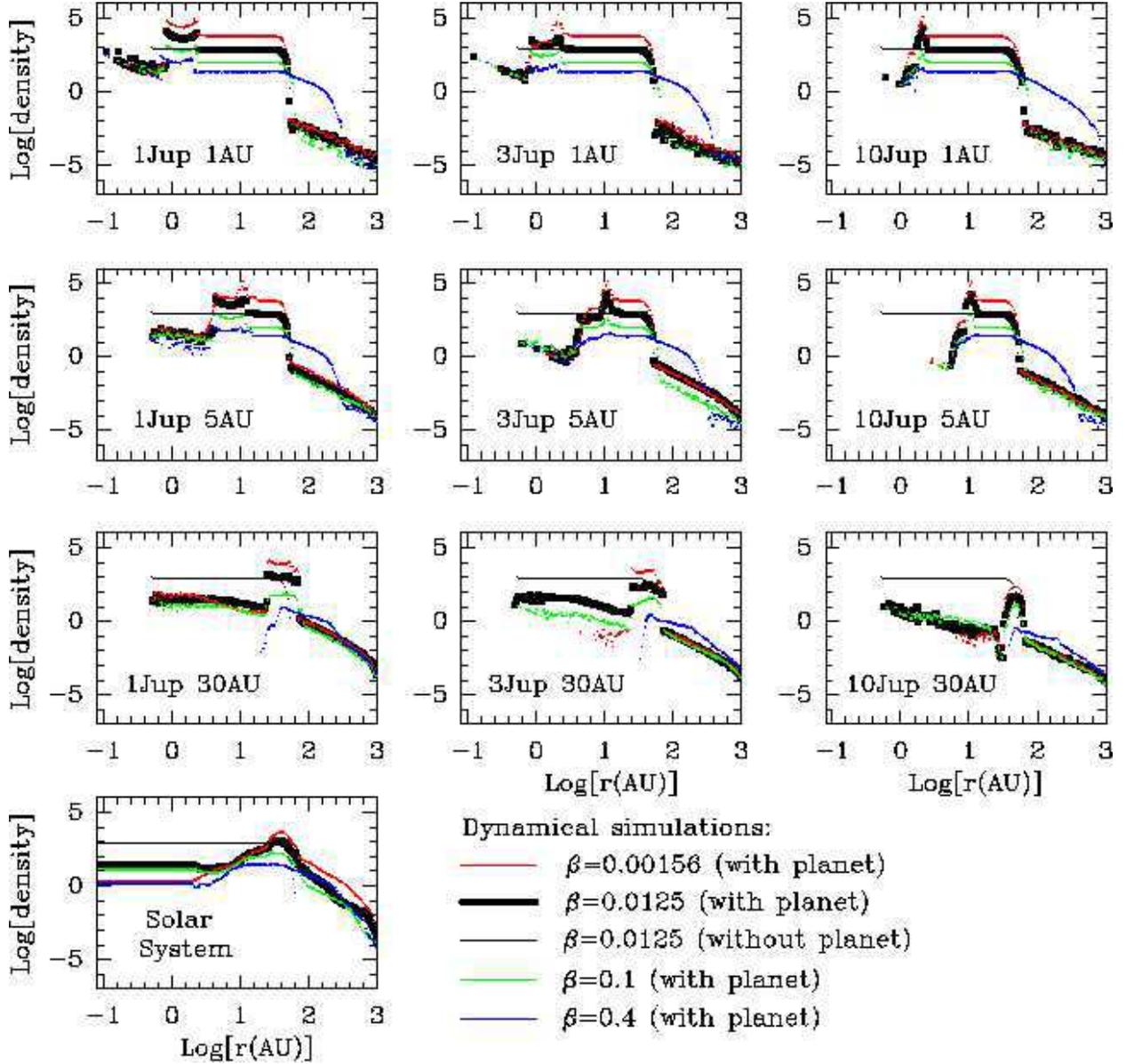}
\caption{Surface density distributions of dust particles with 
four different $\beta$ values, for different planetary systems (indicated in the individual panels).
The units are number of particles per AU$^2$ for a dust production rate of 100 particles
per 1000 years (to be later scaled to the correct dust production rate or total disk
mass). 
$\it{Large~black~dots}$: results from the dynamical simulations when the gravitational 
perturbation of the planet is taken into account, for dust particles with $\beta$=0.0125. 
$\it{Red}$: same as before but for $\beta$=0.00156; 
$\it{Green}$: $\beta$=0.1;
$\it{Blue}$: $\beta$=0.4.
$\it{Small~black~dots}$ (or thin solid line): results from the dynamical simulations 
when no planets are present, for dust particles with $\beta$=0.0125.
The Solar System model is extrapolated down to 0.01 AU assuming a flat surface 
density distribution, expected for a collisionless system with grains in circular orbits.}
\end{figure}

\clearpage 

\begin{figure}
\epsscale{1.0}
\plotone{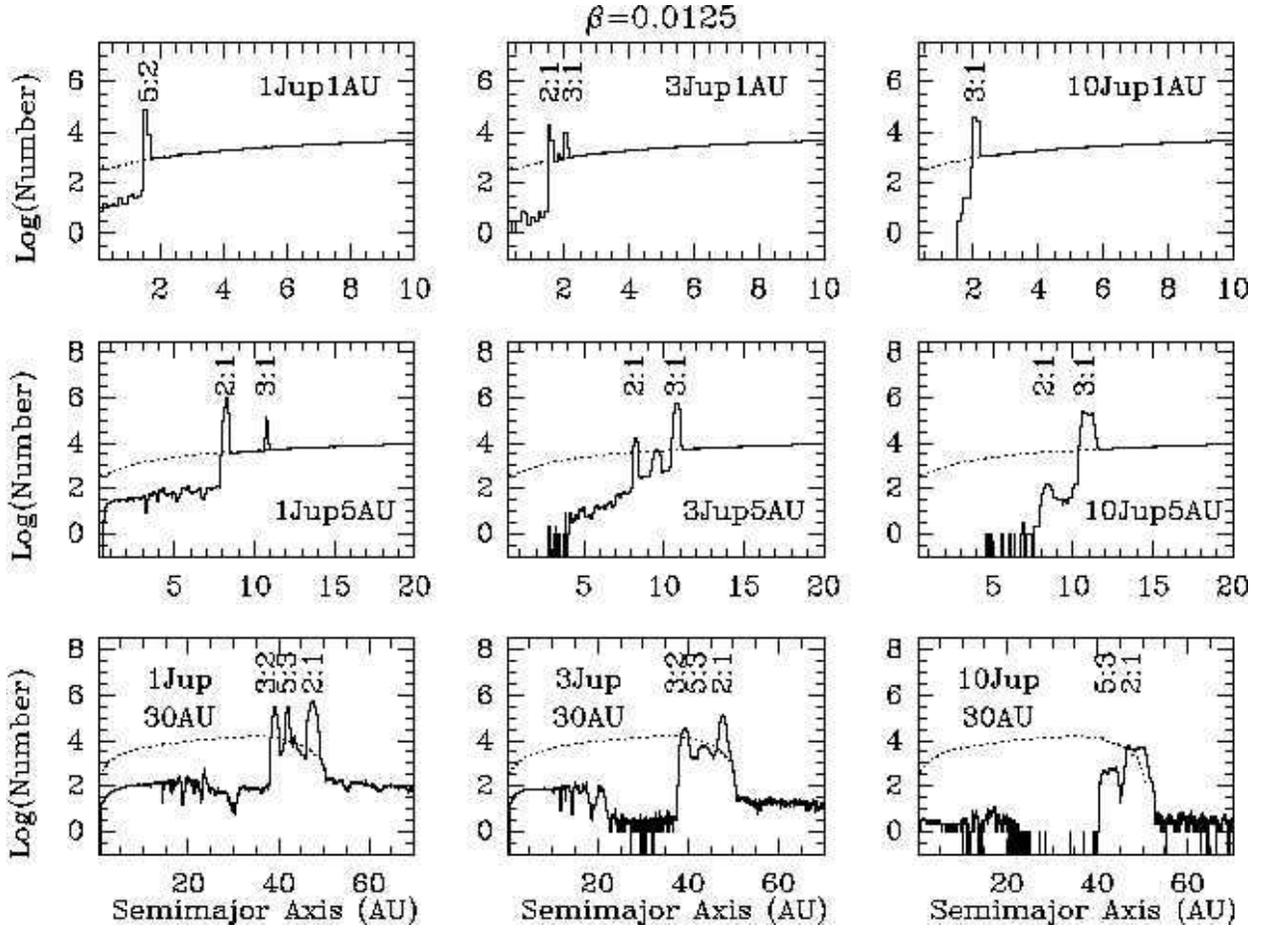}
\caption{``Equilibrium'' semimajor axis distributions in logarithmic scale of the 
dust particles with $\beta$=0.0125, in different single-planet systems
(indicated in the individual panels; $\it{solid~line}$). The 
$\it{dotted~line}$ correspond to a system without planets. 
The trapping of particles in the exterior MMRs with the planet, and 
the depletion of particles inside the planet's orbit are the most prominent 
features in the figure. The y-axis is the number of particles within a given
range of semimajor axis. Only the relative numbers are meaningful.}
\end{figure}

\clearpage 

\begin{figure}
\epsscale{1.0}
\plotone{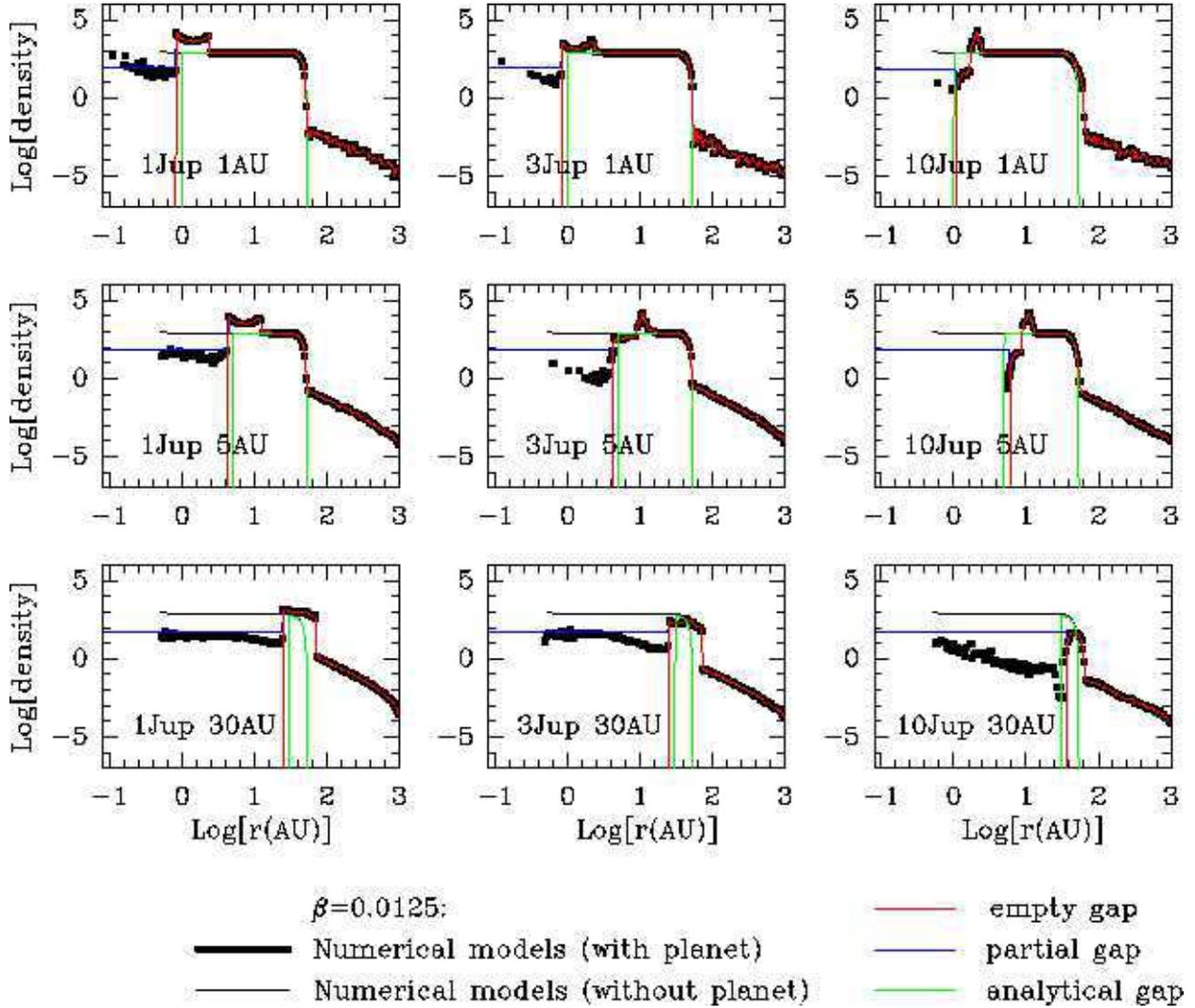}
\caption{Surface density distributions used as input for the radiative transfer code for
dust particles with $\beta$=0.0125, for different planetary systems (indicated in the individual panels).
$\it{Large~black~dots}$: results from the dynamical simulations when a planet is present.
$\it{Small~black~dots}$ (or thin solid line): results from the dynamical simulations 
when no planets are present.
$\it{Red~line}$: surface density distributions with ``empty gap''.
$\it{Blue~line}$: surface density distributions with ``partial gap''.
$\it{Green~line}$: surface density distributions with ``analytical gap''.}
\end{figure}

\clearpage 

\begin{figure}
\plotone{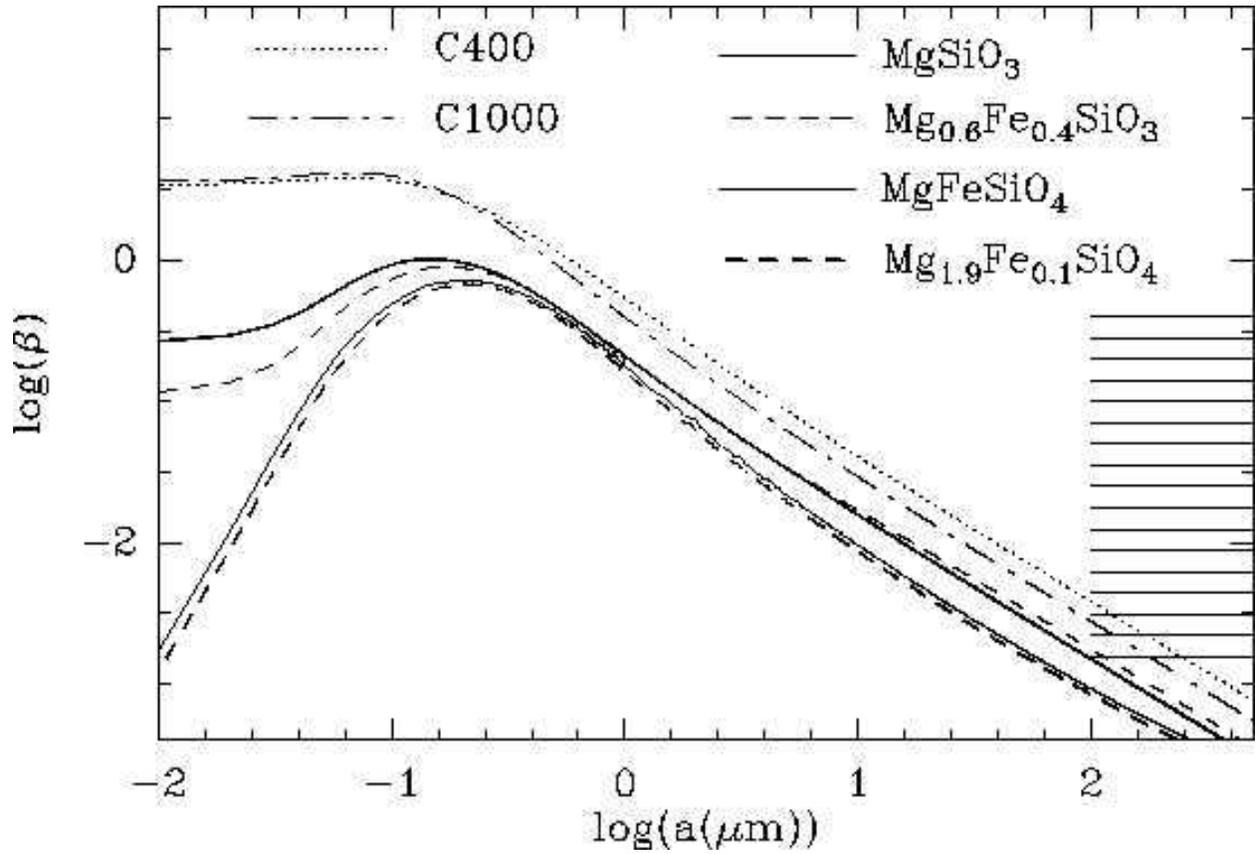}
\caption{Correspondence between $\beta$ and particle radius for 
the grain chemical compositions under consideration. 
The horizontal lines at the far right indicate the 
values of $\beta$ used in the dynamical models.}
\end{figure}

\clearpage 

\begin{figure}
\epsscale{0.35}
\plotone{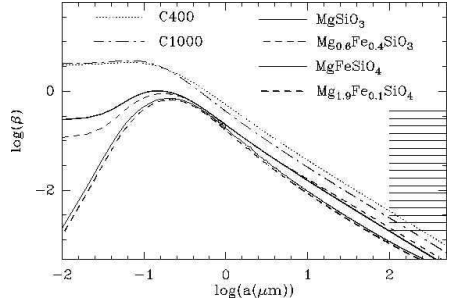}
\caption{SEDs of single particle size disks from the Solar System models. 
Each panel corresponds to a particular grain chemical composition.
Each color corresponds to a different $\beta$ value (or grain size): $\it{Red}$ for
$\beta$=0.4, $\it{Green}$ for $\beta$=0.025 and $\it{Blue}$ for $\beta$=0.00156. 
$\it{Solid~line}$: system with 7 planets; $\it{Dotted~line}$:
system without planets. In all cases the disk is assumed to be at
distance of 50 pc and has a mass of 10$^{-10}$~\msol. The squares
indicate $\it{Spitzer}$ 5--$\sigma$ detection limits.}
\end{figure}

\clearpage 

\begin{figure}
\epsscale{0.75}
\plotone{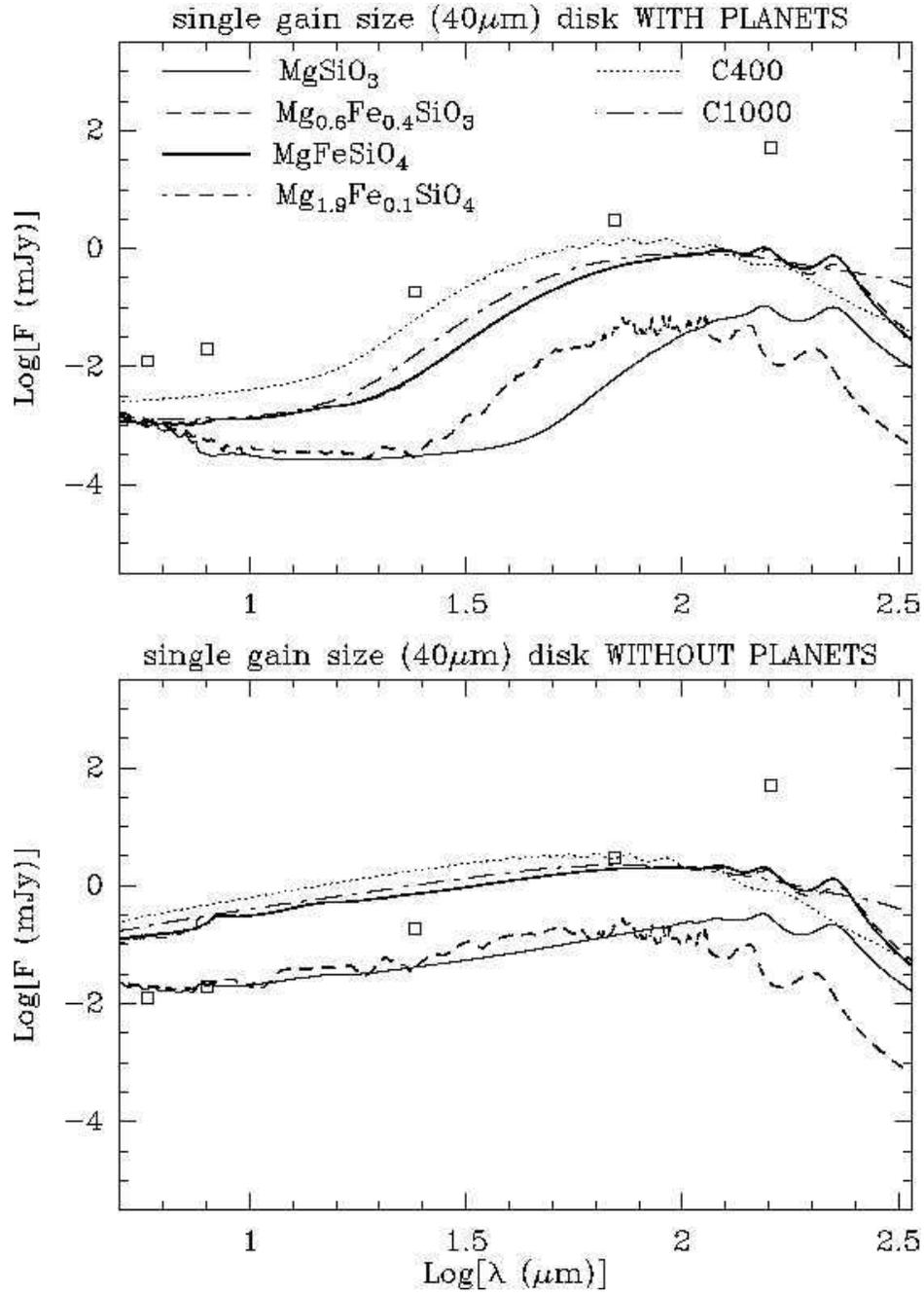}
\caption{SEDs of disks composed of 1 $\mu$m and 40 $\mu$m grains, from the Solar System models, and 
with different grain chemical compositions.}
\end{figure}

\clearpage 

\begin{figure}
\epsscale{0.75}
\plotone{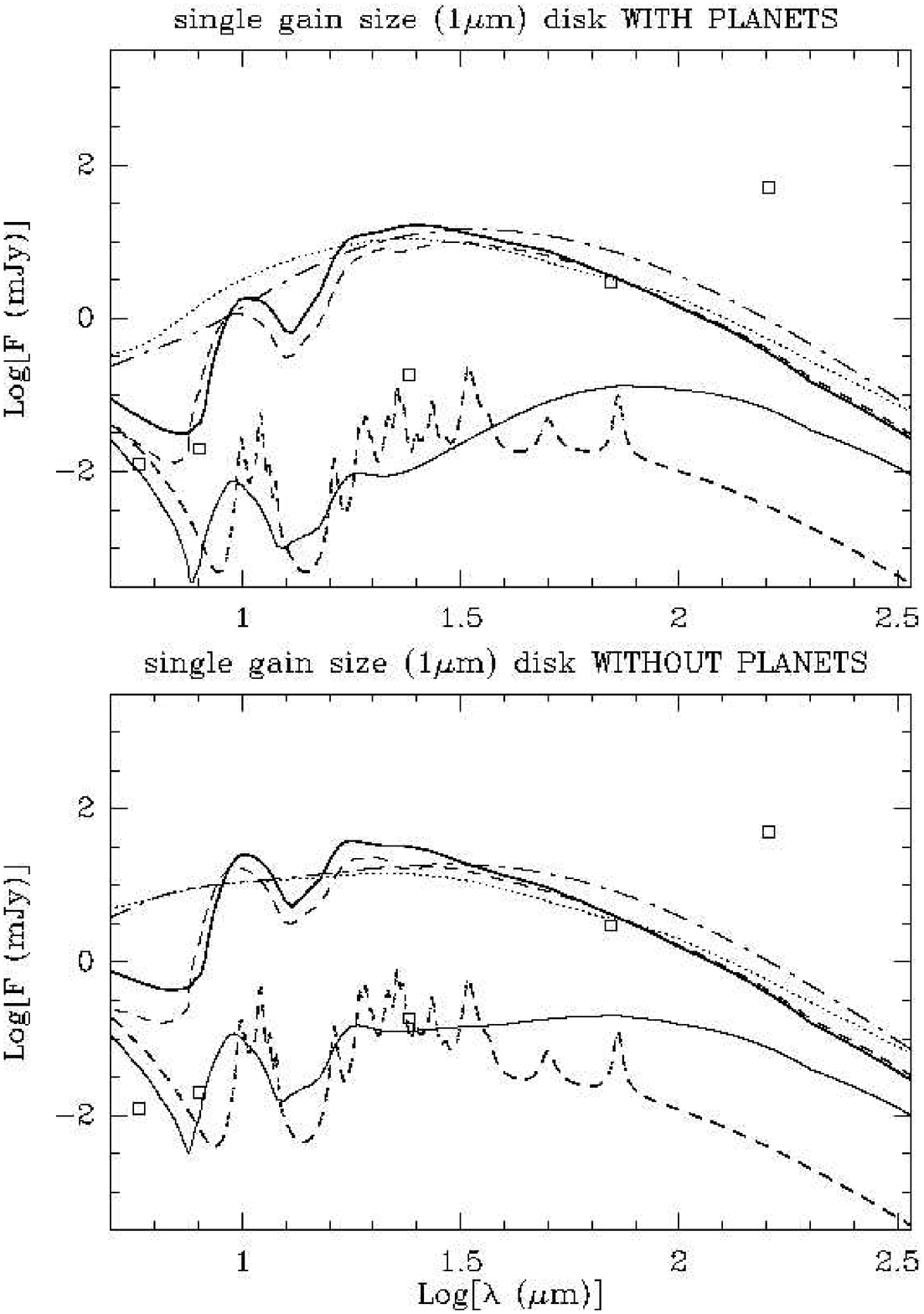}
\end{figure}

\clearpage 

\begin{figure}
\epsscale{0.40}
\plotone{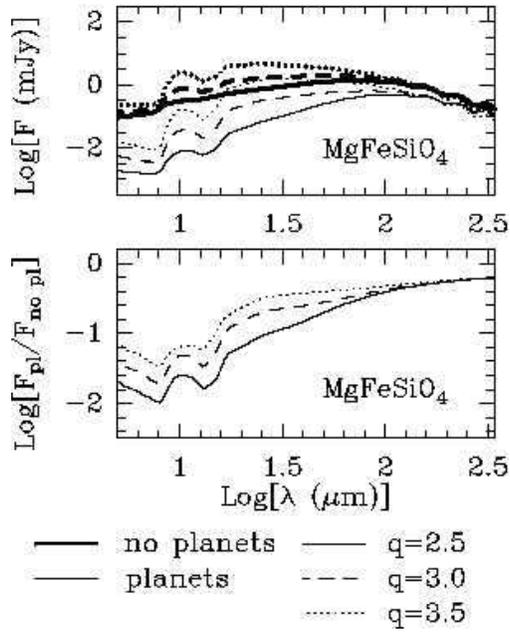}
\caption
{($\it{Top~panel}$) SEDs of dust disks composed of MgFeSiO$_4$ grains, for three particle 
size distribution: $\it {n(b)db=n_0b^{-q}}$, with $\it{q}$=2.5 ($\it{solid~line}$), 
3.0 ($\it{dashed~line}$) and 3.5 ($\it{dotted~line}$).
$\it{Thick~line}$: system without planets; $\it{Thin~line}$:
system with Solar System-like planets. The system is at a 
distance of 50 pc and has a total disk mass of 10$^{-10}$~\msol. 
($\it{Bottom~panel}$) Ratio of the composed SED arising from a system
with planets to that of a system with no planets.}
\end{figure}

\clearpage 

\begin{figure}
\epsscale{1.0}
\plotone{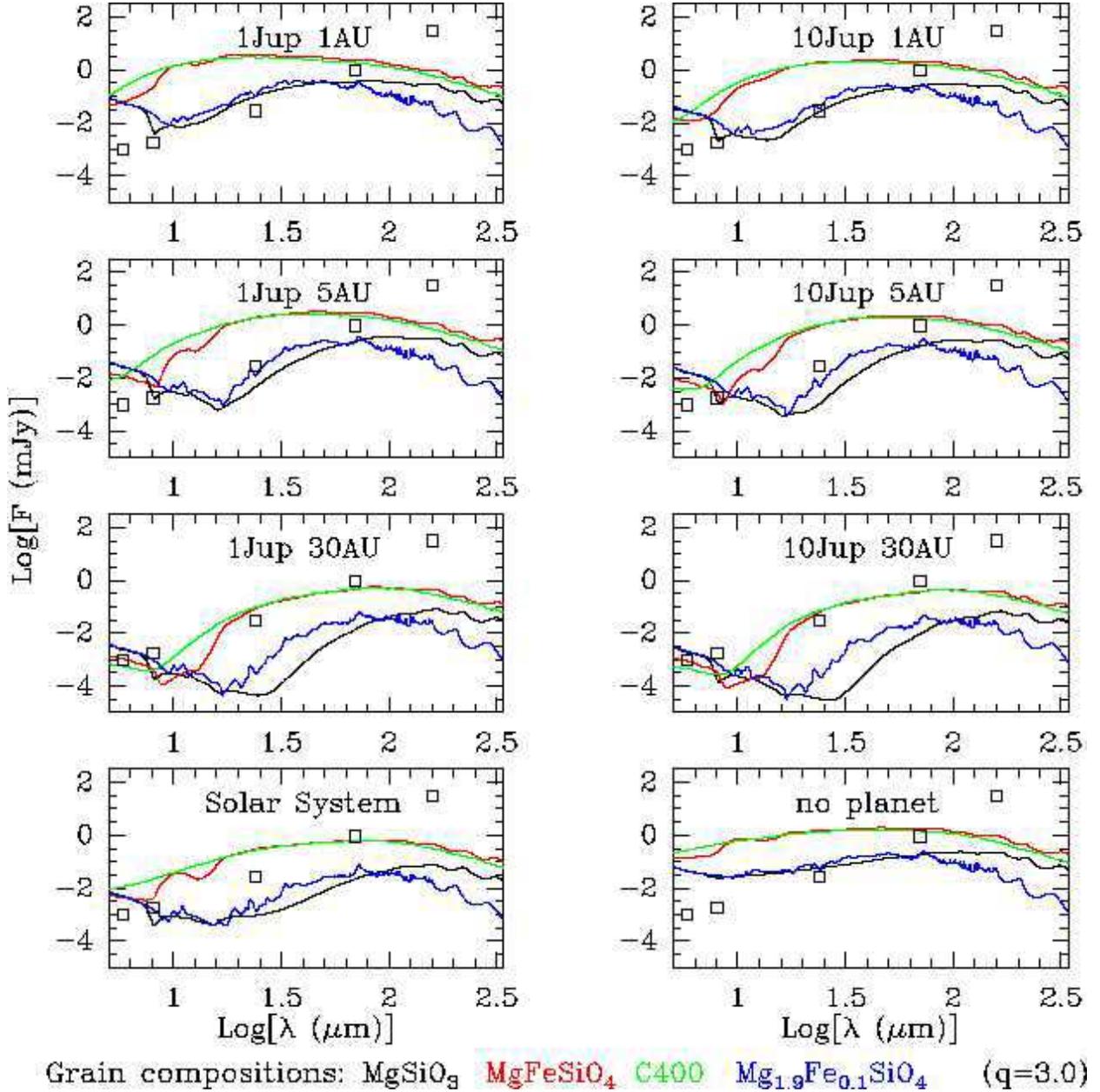}
\caption{SEDs of dust disks in the presence of different planetary 
configurations (indicated at the top of each panel), 
for four grain chemical compositions (in different colors),
and particle size distribution given by $\it {n(b)db=n_0b^{-q}}$ with $\it{q}$=3.0.
In all cases the system is at a distance of 50 pc and has a total disk 
mass of 10$^{-10}$~\msol. The squares indicate $\it{Spitzer}$ 5--$\sigma$ 
detection limits.} 
\end{figure}

\clearpage 

\begin{figure}
\epsscale{1.0}
\plotone{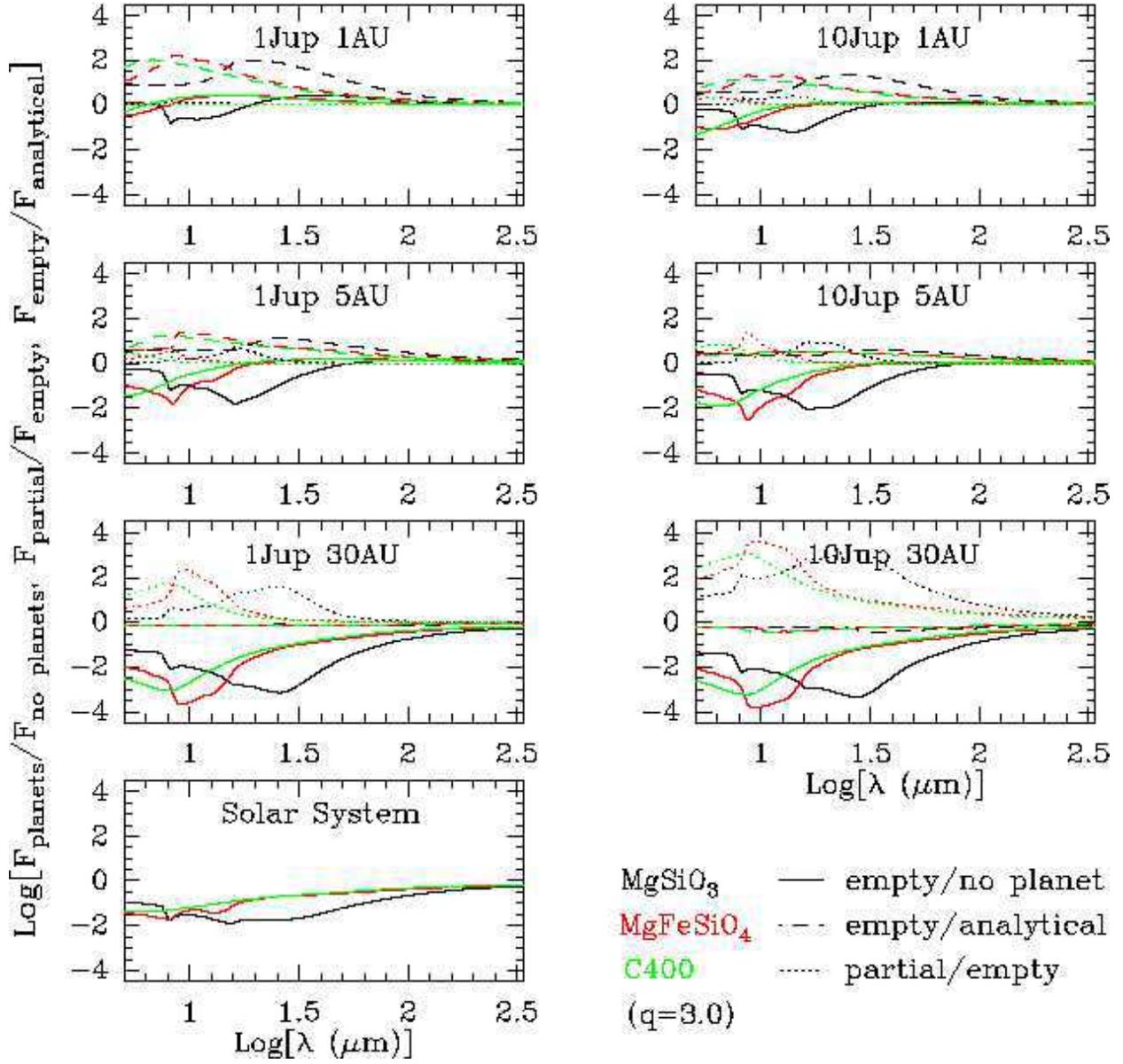}
\caption
{Ratio of the SEDs that arise from different modeled disks:
$\it{Solid~line}$: F$_{planet}$/F$_{no~planet}$ (= F$_{empty~gap}$/F$_{no~planet}$);
$\it{Dotted~line}$: F$_{partial~gap}$/F$_{empty~gap}$; and
$\it{Dashed~line}$: F$_{empty~gap}$/F$_{analytical~gap}$. 
The different colors correspond to different grain chemical compositions. In all 
cases $\it{q}$=3.0.}
\end{figure}

\clearpage 

\begin{figure}
\epsscale{0.7}
\plotone{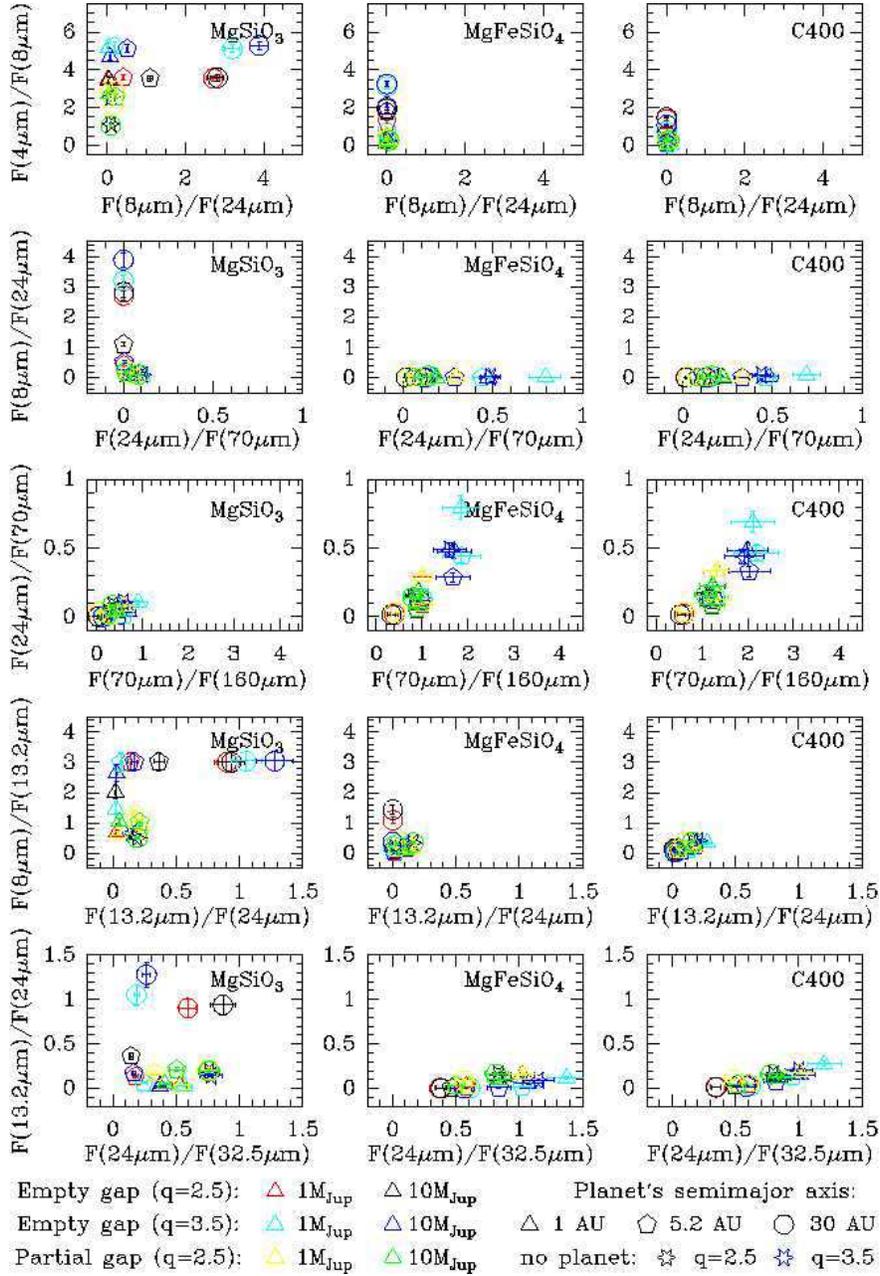}
\caption{Predicted $\it{Spitzer}$ broadband colors for the SEDs in Figure 9.
The different symbols correspond to different planetary systems.
The symbol shape indicates the planet semimajor axis.
The symbol color indicates planet mass, power law index for the particle 
size distribution, and whether the gap is empty or partially filled.
The error bars indicate (optimistic) $\it{Spitzer}$ calibration uncertainties:
3\% for IRAC in all bands; 10\% for IRS; 5\% for MIPS 24 $\mu$m; 
10\% for MIPS 70 $\mu$m; 20\% for MIPS 160 $\mu$m.}
\end{figure}

\clearpage 

\begin{figure}
\epsscale{0.35}
\plotone{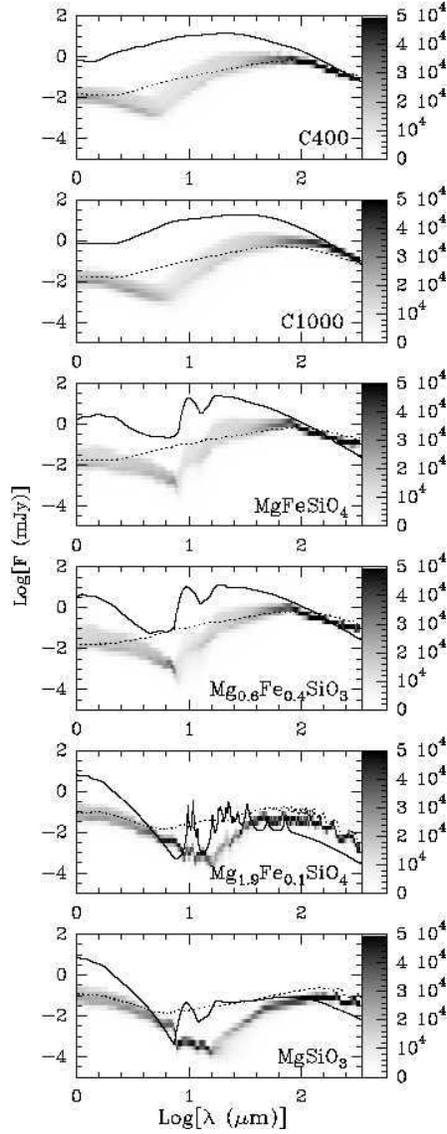}
\caption
{SED probability distributions. The $\it{gray-scale}$ corresponds to SEDs of
disks with embedded planets between 1 AU and 30 AU, and with different dust depletion
factors. The $\it{solid}$ and $\it{dotted}$ lines represent the SEDs from the disks 
without planets for $\beta$-values of 0.4 and 0.00156, respectively. Each panel
corresponds to a different grain mineralogy.}
\end{figure}

\clearpage 

\begin{figure}
\epsscale{0.7}
\plotone{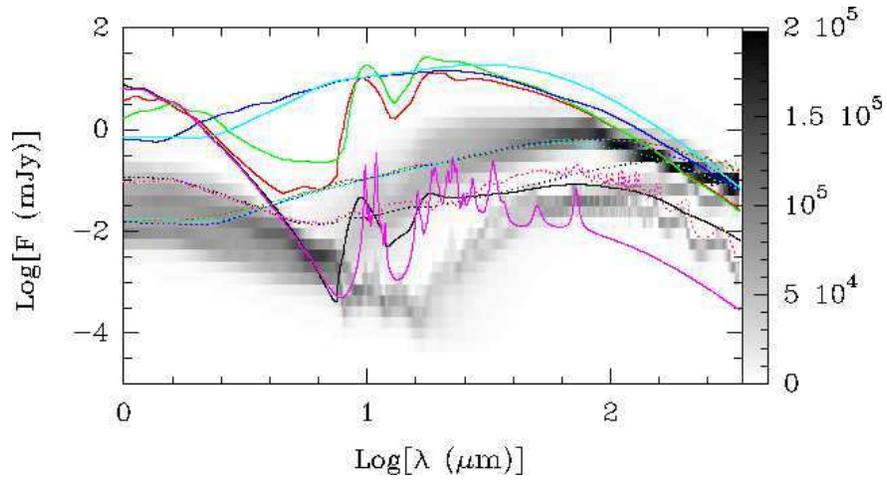}
\caption
{Same as Figure 12, with the six mineralogies all plotted together in gray-scale  
(a total of 21$\times$36$\times$11$\times$6 = 49896 SEDs). The 
SEDs from the disks without planets are shown in the $\it{solid}$ and $\it{dotted}$
lines, with each color corresponding to one mineralogy. 
$\it{black}$: MgSiO$_3$; 
$\it{red}$: Mg$_{0.6}$Fe$_{0.4}$SiO$_3$; 
$\it{green}$: MgFeSiO$_4$;
$\it{magenta}$: Mg$_{1.9}$Fe$_{0.1}$SiO$_4$; 
$\it{dark~blue}$: C400; 
$\it{light~blue}$: C1000.}
\end{figure}

\clearpage 

\begin{figure}
\epsscale{0.7}
\plotone{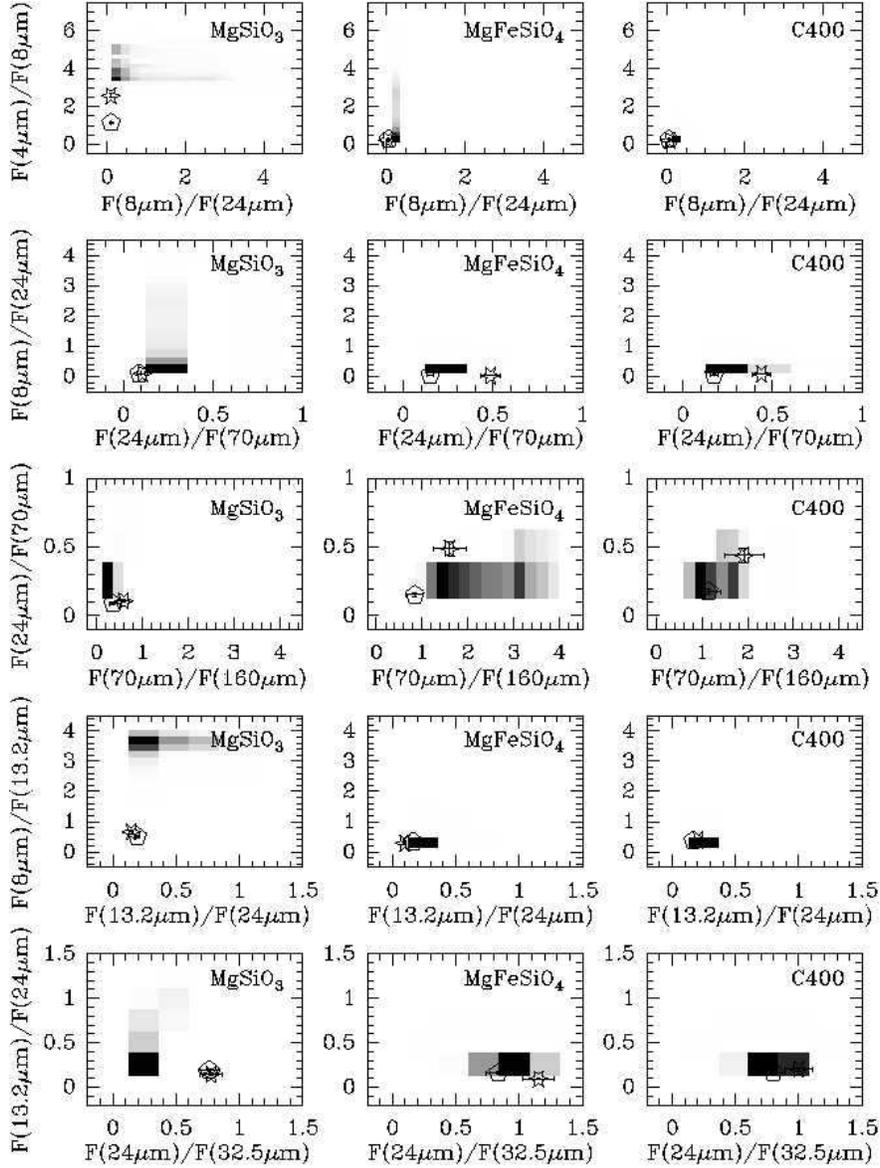}
\caption
{Predicted $\it{Spitzer}$ broadband colors for the SEDs in Figure 12.
The $\it{gray-scale}$ corresponds to the colors from disks with embedded planets 
between 1 AU and 30 AU, and with different dust depletion factors. The symbols
(pentagon and star) correspond to the colors derived from disks without planets
(for power law indexes of 2.5 and 3.5, respectively).
The error bars indicate (optimistic) $\it{Spitzer}$ calibration uncertainties:
3\% for IRAC in all bands; 10\% for IRS; 5\% for MIPS 24 $\mu$m; 
10\% for MIPS 70 $\mu$m; 20\% for MIPS 160 $\mu$m.}
\end{figure}

\clearpage 

\begin{figure}
\epsscale{0.4}
\plotone{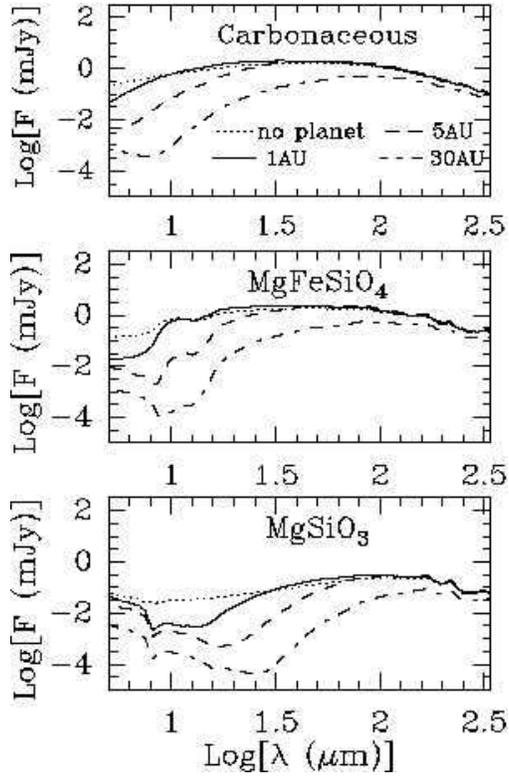}
\caption{SEDs of dust disks in the presence of different planetary 
configurations [$\it{solid}$: 3M$_{Jup}$ at 1 AU; $\it{dashed}$: 3M$_{Jup}$ at 5 AU;
$\it{dashed-dotted}$: 3M$_{Jup}$ at 30 AU; $\it{dotted}$: system without planets]. Results
are shown for three grain chemical compositions (indicated in each panel), and particle size distribution 
given by $\it {n(b)db=n_0b^{-q}}$ with $\it{q}$=3.0.
In all cases the system is at a distance of 50 pc and has a total disk 
mass of 10$^{-10}$~M$\odot$.}
\end{figure}

\clearpage 

\begin{figure}
\epsscale{0.40}
\plotone{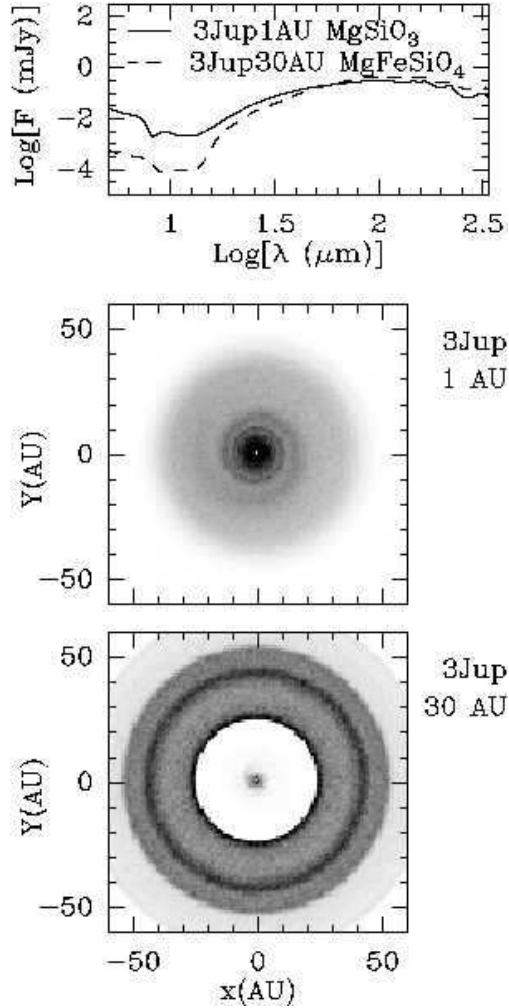}
\caption
{(top) Possible degeneracy between the grain chemical composition and the 
location of the planet clearing the gap. $\it{Solid~line}$: SED of dust disk 
composed of MgSiO$_3$ grains with a 3M$_{Jup}$ planet at 1 AU; 
$\it{dashed~line}$: same for MgFeSiO$_4$ grains with a
3M$_{Jup}$ planet at 30 AU. In both cases $\it{q}$=2.5. (middle) and (bottom)
Brightness density distributions at 70 $\mu$m (assuming graybody emission from 
12 $\mu$m grains) expected from a disk with a 3M$_{Jup}$ planet at 1 AU and 30 AU, 
respectively (shown in arbitrary units). High resolution images are needed 
to solve the degeneracy.}
\end{figure}


\begin{thebibliography}{}
\bibitem[Backman \& Paresce~(1993)]{back93}
Backman, D.E. \& Paresce, F. 1993, Protostars and Planets III 
(ed. E.H. Levy \& J.I Lunine, Tucson: Univ. of Arizona Press), 1253
\bibitem[Beichman et al.~(2005)]{beic05}
Beichman C.A., Bryden, G., Rieke, G.H. et al., in preparation, 2005
\bibitem[Briggs~(1962)]{brig62}
Briggs, R.E. 1962, AJ, 67, 710
\bibitem[Burns, Lamy \& Soter~(1979)]{burn79}
Burns, J.A., Lamy, P.L., Soter, S. 1979, Icarus, 40, 1
\bibitem[Dent et al.~(2000)]{dent00}
Dent, W.R.F., Walker, H.J., Holland, W.S. \& Greaves, J.S. 2000, 
MNRAS, 314, 702
\bibitem[Dorschner et al.~(1995)]{dors95}
Dorschner, J., Begemann, B., Henning, Th., J\"{a}ger, C. \& 
Mutschke, H. 1995, A\&A, 300, 503
\bibitem[Draine \& Lee~(1984)]{drai84}
Draine, B.T. \& Lee, H.M. 1984, ApJ, 285, 89 
\bibitem[Gorlova et al.~(2004)]{gorl04}
Gorlova, N., Padgett, D.L., Rieke, G.H. et al. 2004, ApJS, 154, 448 
\bibitem[Greaves et al.~(1998)]{grea98}
Greaves, J.S., Holland, W.S., Moriarty-Schieven, G. et al. 1998, ApJ, 506, 133
\bibitem[Greaves, Mannings \& Holland~(2000)]{grea00}
Greaves, J.S., Mannings, V. \& Holland, W.S. 2000, Icarus, 143, 155
\bibitem[Greaves \& Wyatt~(2003)]{grea03}
Greaves, J.S. \& Wyatt, M.C. 2003, MNRAS, 345, 1212
\bibitem[Greaves et al.~(2004)]{grea04}
Greaves, J.S., Holland, W.S., Jayawardhana, R., Wyatt, M.C. \& Dent,W.R.F 2004, MNRAS, 348, 1097
\bibitem[Habing et al. (2001)]{habi01}
Habing, H.J., Dominik, C., Jourdain de Muizon, M. et al. 2001, A\&A, 365, 545
\bibitem[Henning et al.~(1995)]{henn95}
Henning, Th., Begemann, B., Mutschke, H., \& Dorschner, J. 1995, A\&AS, 112, 143 
\bibitem[Henning \& Stognienko~(1996)]{henn96}
Henning, Th., \& Stognienko, R. 1996, A\&A, 311, 291
\bibitem[Holland et al. (2003)]{holl03}
Holland, W.S. et al. 2003, ApJ, 582, 1141
\bibitem[J\"{a}ger et al. (1998)]{jage98}
J\"{a}ger, C., Mutschke, H. \& Henning, Th. 1998, A\&A, 332, 291
\bibitem[Kuchner \& Holman~(2003)]{kuch03}
Kuchner, M.J., Holman, M.J. 2003, ApJ, 588, 110
\bibitem[Labs \& Neckel~(1968)]{labs68}
Labs, D. \& Neckel, H. 1968, ZA, 69, 1 
\bibitem[Landgraf et al.~(2002)]{land02}
Landgraf, M., Liou, J-C., Zook \& H.A, Gr\"{u}n, E. 2002, AJ, 123, 2857
\bibitem[Liou, Zook \& Dermott (1996)]{liou96}
Liou, J.C., Zook, H.A. \& Dermott, S.F. 1996, Icarus, 124, 429
\bibitem[Liou \& Zook (1999)]{liou99}
Liou, J.C. \& Zook, H.A., 1999, AJ, 118, 580
\bibitem[Malfait et al.~(2000)]{malf00}
Malfait, K., Waelkens, C., Bouwman, J., De Koter, A. \& Waters, L.B.F.M. 1999, A\&A, 345, 181 
\bibitem[Malhotra et al. (2000)]{malh00}
Malhotra, R., Duncan, M.J., Levison, H.F. 2000, Protostars and Planets IV 
(eds Mannings, V., Boss, A.P., Russell, S. S.), 1231
%\bibitem[Marcy et al~(2003)]{marc03}
%Marcy, G.W., Butler, R.P., Fischer, D.A. \& Vogt, S.S. 2003, Scientific Frontiers in Research on Extrasolar Planets (eds Deming \& Seager) ASP Conference Series, 294, 1
\bibitem[Moro-Mart\'{\i}n \& Malhotra (2002)]{ama02}
Moro-Mart\'{\i}n, A. \& Malhotra, R. 2002, AJ, 124, 2305
\bibitem[Moro-Mart\'{\i}n \& Malhotra (2003)]{ama03}
Moro-Mart\'{\i}n, A. \& Malhotra, R. 2003, AJ, 125, 2255
\bibitem[Mouillet et al~(1997)]{moui97}
Mouillet, D., Larwood, J.D., Papaloizou, J.C.B. \& Lagrange, A.M. 1997, MNRAS, 292, 896
\bibitem[Ozernoy et al.~(2000)]{ozer00}
Ozernoy, L.M., Gorkavyi, N.N., Mather, J.C. \& Taidakova, T.A. 2000, ApJ, 537, 147
\bibitem[Pantin, Lagage \& Artymowicz (1997)]{pant97}
Pantin, E., Lagage, P.O. \& Artymowicz, P. 1997, A\&A, 327, 1123
\bibitem[Quillen \& Thorndike~(2002)]{quil02}
Quillen, A.C. \& Thorndike, S. 2002, ApJL, 2002, 578, 149
\bibitem[Reach et al. (2003)]{reac03}
Reach, W.T., Morris, P., Boulanger, F. \& Okumura, K. 2003, Iracus, 164, 384
\bibitem[Rieke et al. (2005)]{riek05}
Rieke et al. 2004, in press
\bibitem[Roques et al. (1994)]{roqu94}
Roques, F., Scholl, H., Sicardy, B. \& Smith, B.A. 1994, Icarus, 108, 37
\bibitem[Savage \& Mathis~(1979)]{sava79}
Savage, B.D. \& Mathis, J.S. 1979, ARA\&A, 17, 73 
\bibitem[Spangler et al (2001)]{span01}
Spangler, C., Sargent, A.I. \& Silverstone, M.D. 2001, ApJ, 555, 932
\bibitem[Weinberger et al. (2003)]{wein03}
Weinberger, A.J., Becklin, E.E. \& Zuckerman, B. 2003, ApJ, 584, L33
\bibitem[Wilner et al. (2002)]{wiln02}
Wilner, D.J, Holman, M.J., Kuchner, M.J. \& Ho, P.T.P 2002, ApJ, 569, L115
\bibitem[Wolf \& Hillenbrand (2003)]{wolf03}
Wolf, S. \& Hillenbrand, L.A. 2003, ApJ, 596, 603
\bibitem[Warren~(1984)]{warr84}
Warren, S. G. 1984, Appl. Opt., 23, 1206 
\bibitem[Wyatt, Dermott \& Telesco~(1999)]{wyat99}
Wyatt, M.C., Dermott, S.F. \& Telesco, C.M., 1999, ApJ, 527, 918
\bibitem[Wyatt \& Dent (2002)]{wyat02}
Wyatt, M.C. \& Dent, W.R.F. 2002, MNRAS, 334, 589
%\bibitem[Zuckerman~(2001)]{zuck01}
%Zuckerman, B. 2001, ARA\&A, 39, 549
\end{thebibliography}
\end{document}